\documentclass[]{interact}

\usepackage{epstopdf}
\usepackage[caption=false]{subfig}

\usepackage[numbers,sort&compress]{natbib}
\bibpunct[, ]{[}{]}{,}{n}{,}{,}
\makeatletter
\def\NAT@def@citea{\def\@citea{\NAT@separator}}
\makeatother

\theoremstyle{plain}

\theoremstyle{definition}

\theoremstyle{remark}

\begin{document}

\articletype{}

\title{Femtosecond laser inscriptions in Kerr nonlinear transparent media: dynamics in the presence of K-photon absorptions, radiative recombinations and electron diffusions}

\author{
\name{Emmanuel O. Akeweje\textsuperscript{a}, G. Bader\textsuperscript{b}, Alain M. Dikand\'e\textsuperscript{c}\thanks{CONTACT A.~M. Dikand\'e. Email: dikande.alain@ubuea.cm} and P. Kameni Nteutse\textsuperscript{d}}
\affil{\textsuperscript{a}African Institute for Mathematical Sciences (AIMS-Ghana), East Legon Hills, Accra, Ghana.; \\ \textsuperscript{b}Institute of Mathematics, Brandenburg Technical University, Cottbus-Senftenberg, Germany; \\ \textsuperscript{c}Laboratory of Research on Advanced Materials and Nonlinear Science (LaRAMaNS), Department of Physics, Faculty of Science, University of Buea P.O. Box 63 Buea, Cameroon; \\ \textsuperscript{c}African Institute for Mathematical Sciences (AIMS-Cameroon), Crystal Garden P.O. Box 608 Limbe, Cameroon.}
}

\maketitle

\begin{abstract}
Femtosecond lasers interacting with Kerr nonlinear optical materials, propagate in form of filaments due to the balance of beam diffraction by self-focusing induced by the Kerr nonlinearity. Femtosecond laser filamentation is a universal phenomenon that belongs to a general class of processes proper to ultrashort lasers processing systems, associated with the competition between nonlinearity and dispersion also known to promote optical solitons. The present work considers a model describing femtosecond laser inscriptions in a transparent medium with Kerr nonlinearity. Upon inscription, the laser stores energy in the optical material which induces an electron plasma. The model consists of a cubic complex Ginzburg-Landau equation, in which an additional $K$-order nonlinear term takes into account $K$-photon absorption processes. The complex Ginzburg-Landau equation is coupled to a time first-order nonlinear ordinary differential equation, accounting for time evolution of the plasma density. The main objective of the study is to examine effects of the competition between multi-photon absorptions, radiative recombination and electron diffusion processes, on temporal profiles of the laser amplitude as well as of the plasma density. From numerical simulations, it is found that when the photon number (i.e. $K$) contributing to multiphoton ionization is large enough, taking the electron diffusion processes into account favours periodic structures in temporal profiles both of the laser and the plasma density. The pulse repetition rate in the optical soliton train is increased with increase of the electron diffusion coefficient, while the plasma density is a train of multi-periodic anharmonic wave patterns.  
\end{abstract}


\section{Introduction}
\label{intro}
Femtosecond lasers have become one of most efficient tools in a large variety of modern industrial material processings \cite{r1,r2,r3,r4}. Their applications extend from accurate manufacturing of electronic devices to fine cutting, polishing and machining of hard metals, ceramics and soft plastics for fabrications of various micro textures with fine and stylished designs \cite{r2,r3,r4,r5}. These applications are the fruit of outstanding progress in femtosecond laser micromachining technology observed over the recent past \cite{r1,r2,r3,r4,r5,r5a}, that motivated a renewed interest in ultrashort laser interactions with dielectric media \cite{r1,r2,r3,r4,r5,r5a}.
\par In particular, the interaction of femtosecond lasers with a transparent material is a strongly nonlinear process that leads to a highly localized energy storage in the optical material \cite{r6,r7,r8,r9,r10,r11}. In this nonlinear process two physical phenomena are crucial for an optimum energy storage, namely multiphoton ionizations and inverse Bremsstrahlung absorption \cite{r9}. When the first free electrons, produced via multiphoton ionization processes, accumulate sufficient energy by inverse Bremsstrahlung absorption, they trigger avalanche ionization that causes the generation of an electron plasma \cite{r9}. This electron plasma further absorbs energy from the incident laser field to move toward the incoming laser field, thus forming new layers of electron plasma in the material \cite{r9}. The generation of electron plasma can be accompanied with an increase of the absorption coefficient of the optical medium, favoring energy deposition from the radiation field at a relatively short time scale \cite{r9}. \par
Theoretical investigations of femtosecond laser interaction with transparent media have been carried out in some past studies \cite{r6,r7,r8,r9,r10,r11}. In these studies the optical field propagation in the transparent medium is described by the complex Ginzburg-Landau equation, while time evolution of the plasma density is represented by a Drude-type equation, with specific terms accounting for physical ingredients assumed to contribute to photoionization processes. In principle, these models are intended to give account of the laser self-focusing due to Kerr nonlinearity of the transparent medium, enhanced by multiphoton absorption processses \cite{r11a,r11b}. However most of these past studies \cite{r6,r7,r8,r9,r10,r11} were focused mainly on assessing thermal effects related to material processings using femtosecond lasers, and particularly their impact on the quality (e.g. the degree of fineness and refinement) of processed materials. Nevertheless in relatively more recent works \cite{peg1,peg2}, attention was paid on dynamical properties of some among the models proposed e.g. in refs. \cite{r6,r7,r8,r9,r10,r11}. Thus in ref. \cite{peg1} the laser dynamics, together with time evolution of the plasma density, were investigated considering the model proposed by Petrovic et al. \cite{r11}, according to which the plasma generation was governed essentially by avalanche ionization and K-photon absorption processes. From the assumption that depending on conditions in the optical medium, the femtosecond laser will operate in one of two distinct regimes i.e. either continuous-wave or soliton regimes, it was established \cite{peg1} that for this specific model a decrease in $K$, the number of simultaneously absorbed photons, would stabilize continuous waves in the anomalous
dispersion regime. However any continuous wave propagating in the transparent medium in the normal dispersion regime, is expected to be always unstable irrespective of the value of $K$. The same study \cite{peg1} also established that large values of the phase shift per laser roundtrip, will favor a stronger stability of continuous-wave operation in the normal dispersion regime for any value of $K$, whereas in the anomalous dispersion regime only for small values of $K$ continuous waves would be stable. Extending this first study, in ref. \cite{peg2} the same authors considered the problem but including electron-hole radiative recombination processes \cite{r9}. In this second work they found that in steady state, the electron plasma density will strongly depend on the electron-hole radiative recombination coefficient. They obtained that an increase of the electron-hole radiative recombination coefficient causes the laser amplitude to increase in the soliton regime, whereas the electron plasma density is more and more strongly decreased in time as the radiative recombination coefficient is increased. \par
 In the studies of refs. \cite{peg1,peg2} however, dynamical properties of femtosecond lasers in optical Kerr media were investigated ignoring elctron diffusion processes. Yet in several studies and particularly the work of ref. \cite{r9}, reviewing recent experimental and theoretical developments on the topic, the role of electron diffusions in the buildup of the electron plasma was unambiguously emphasized. In fact, taking into consideration the contribution of electron diffusion processes in the generation of the electron plasma, could be relevant for a better description of dynamical properties of the optical field and the plasma density. \par
 In this work we investigate the dynamics of femtosecond lasers interacting with transparent material with optical Kerr nonlinearity, in the presence of avalanche ionizations, electron-hole radiative recombinations and electron diffusion processes \cite{r9}. Since in two previous studies, we considered the same problem restricting analysis to avalanche ionizations \cite{peg1}, and to a combination of avalanche ionizations and radiative recombination processes \cite{peg2} as being responsible for the electron plasma generation, in the present study our attention will be focused on the influence of electron diffusion processes on dynamical properties of the laser and the plasma density. Starting we look at characteristic properties of the model in the steady-state regime. This first analysis will provide relevant insight onto the influence of electron diffusions on fixed-point solutions to the model equations. Then temporal profiles of the laser amplitude and the plasma density will be generated by solving numerically the equations of motion. In this second analysis, emphasis will be on the effects of varying the electron diffusion coefficient on temporal profiles of the laser amplitude and the plasma density. For this dynamical analysis numerical solutions will be sought in two distinct regimes, namely the regime of very small diffusion coefficient and the regime of relatively large diffusion coefficient.  
 
\section{The model}
\label{two}
 Femtosecond laser interaction with transparent media is a nonlinear process that involves an ultrashort laser field (of the order of femtosecond), focused on an optical material with a nonlinear refractive index. Upon absorption, the laser field propagates in the optical material storing heat along its path, thus ionizing the material and releasing electrons via inverse Bremsstrahlung absorption. Once released free electrons contribute to the avalanche ionization by favoring multiphoton recombination processes, and their diffusions in the nonlinear optical medium.\par Couairon and Mysyrowicz \cite{r9} suggested that in the case of uni-directional propagation, the dynamics of a femtosecond laser and of the density of the electron plasma, could be represented by the following set of coupled nonlinear equations: 
\begin{eqnarray}
	i\frac{\partial u}{\partial z}&=&\delta\frac{\partial^2u}{\partial t^2}-\sigma|u|^2u-i\gamma(1-i\omega_0\tau_0)\rho u-i\mu|u|^{2K-2}u, \label{j1}\\
	\frac{\partial \rho}{\partial t}&=&\nu |u|^2 \rho + \alpha |u|^{2K}-\eta \rho - a\rho^{2} . \label{j2}
\end{eqnarray}
More explicitely eq. (\ref{j1}) describes a complex field $u(z,t)$ propagating along $z$, where the first term in the right-hand side accounts for the group-velocity dispersion, the second term describes the Kerr nonlinearity of the transparent medium, the third term describes both the plasma absorption and laser defocusing, and the last term accounts for multiphoton absorption processes. The second equation, i.e. eq. (\ref{j2}), governs time evolution of the electron plasma density $\rho$. In this equation, the first term in the right-hand side represents the avalanche impact ionization, the second term describes multiphoton ionization processes, the third term accounts for electron diffusion processes and the fourth term represents electron-hole radiative recombination processes. Characteristic parameters in eqs. (\ref{j1}) and (\ref{j2}) are defined as follow (their units are also found in refs. \cite{r9,peg2}):
\begin{itemize}
\item $K$ is the photon number contributing to multiphoton absorption processes,
 \item $\delta$ is the coefficient of group-velocity dispersion,
 \item $\sigma$ is the Kerr nonlinearity coefficient,
 \item $\omega_0$ (in MHz) and $\tau_0$ (in fs) are the frequency and characteristic relaxation time of the electron plasma respectively,
 \item $\mu$ is the $K$-photon ionization coefficient,
 \item $\nu$ is the coefficient of avalanche impact ionization,
 \item $\alpha$ is the coefficient of plasma balance due to multiphoton absorption,
 \item $\eta$ is the electron diffusion coefficient,
 \item $a$ is the electron-hole radiative recombination coefficient.
\end{itemize}
Experimental values of these characteristic parameters are given in most works where the models are proposed from experiments, and also in the review paper ref. \cite{r9}. Here we are interested in a qualitative analysis of the system dynamics, reason why we will not bother about exact values of these characteristic parameters. However our analysis will take care of their appropriate respective signs. \par 
We are interested in solutions to eq. (\ref{j1}) that describe waves of real amplitude, modulated in space and time according to the following ansatz \cite{peg2,soto,dik1,dik2,dik3}:
\begin{equation}
u(z,t)=A(z,t)\exp i\left[\phi(z,t)-\omega z \right],\label{u} 
\end{equation}
where $A(z,t)$ is the real amplitude, $\phi(z,t)$ is the instantaneous phase also assumed real, and $\omega$ is a nonlinear shift in the propagation constant. Let us introduce a reduced time as \cite{peg2,dik1,dik2,dik3} $\tau=t-\vartheta z$, where $\vartheta$ is the inverse pulse velocity. Using this new time variable we can write $A(z,t)\equiv A(\tau)$ and $\phi(z,t)\equiv \phi(\tau)$. Substituting eq. (\ref{u}) in eqs. (\ref{j1}) and (\ref{j2}) and separating real from imaginary parts, we obtain the following set of coupled first-order nonlinear ordinary differential equations:
\begin{eqnarray}
M' &=& -\frac{\vartheta y}{\delta A} + \frac{\gamma\rho}{\delta}  + \frac{\mu A^{2(K-1)}}{\delta} - \frac{2My}{A}, \label{frs1}\\
y' &=&	\frac{M A\vartheta}{\delta} + \frac{\omega A}{\delta} + \frac{\gamma\omega_o\tau_o\rho A}{\delta} + \frac{\delta A M^2}{\delta} + \frac{\sigma A^3}{\delta}, \label{frs2}\\
\rho' &=& \nu A^2 \rho + \alpha A^{2K} -\eta \rho - a\rho^2, \label{frs3} \\
A'&=&y, \hskip 0.3truecm \phi'=M, \label{frs4}
\end{eqnarray}
 where prime symbols denote first-order derivatives with respect to $\tau$. In our study we shall be intersted mainly in the stationary solutions for which $\vartheta=0$. \cite{peg2,soto}
 
 \section{Fixed-point solutions}
 \label{three}
\begin{figure}\centering
\begin{minipage}{0.48\textwidth}
\includegraphics[width=3.in,height=2.in]{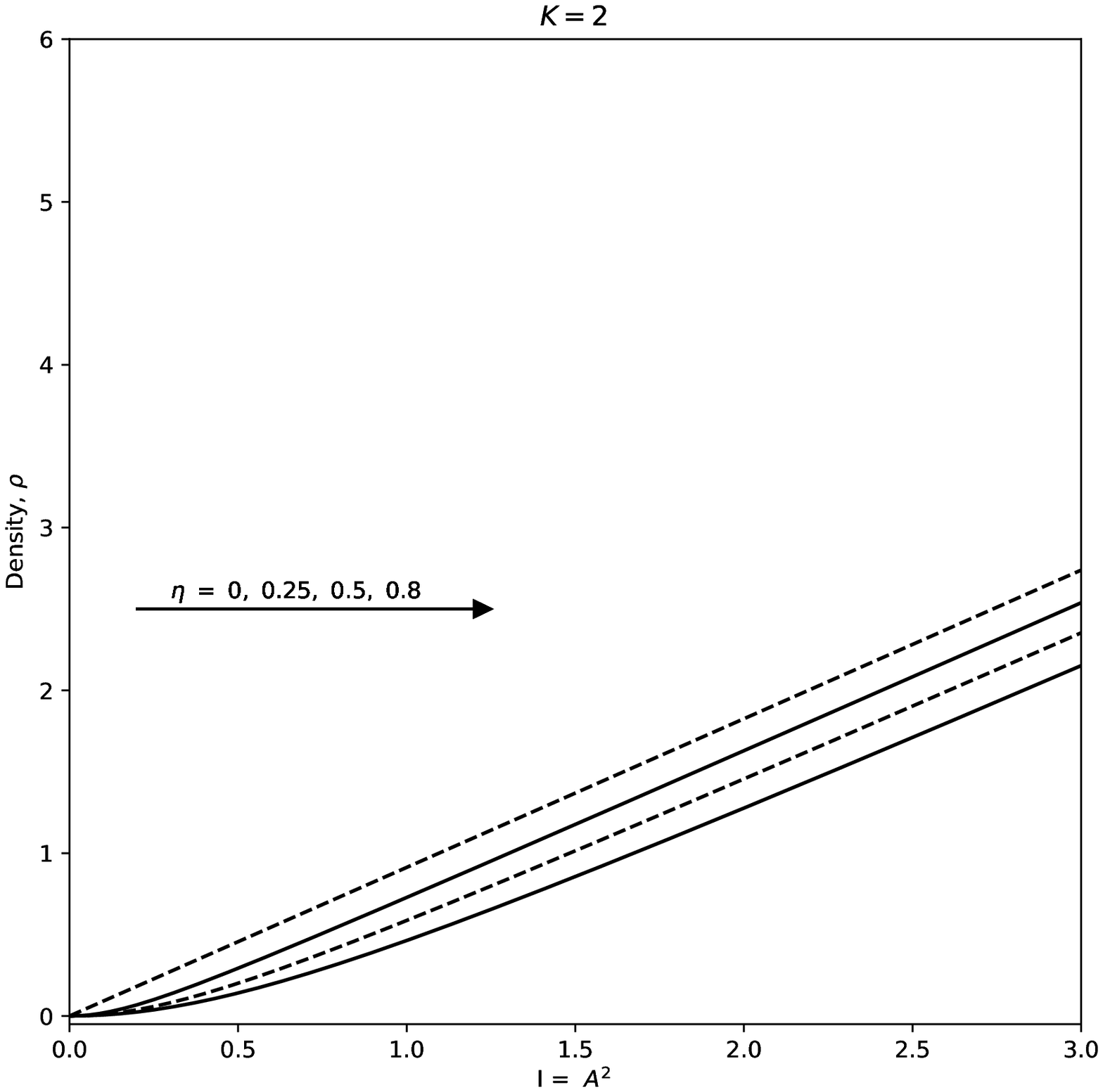}
\end{minipage}%
\begin{minipage}{0.48\textwidth}
\includegraphics[width=3.in,height=2.in]{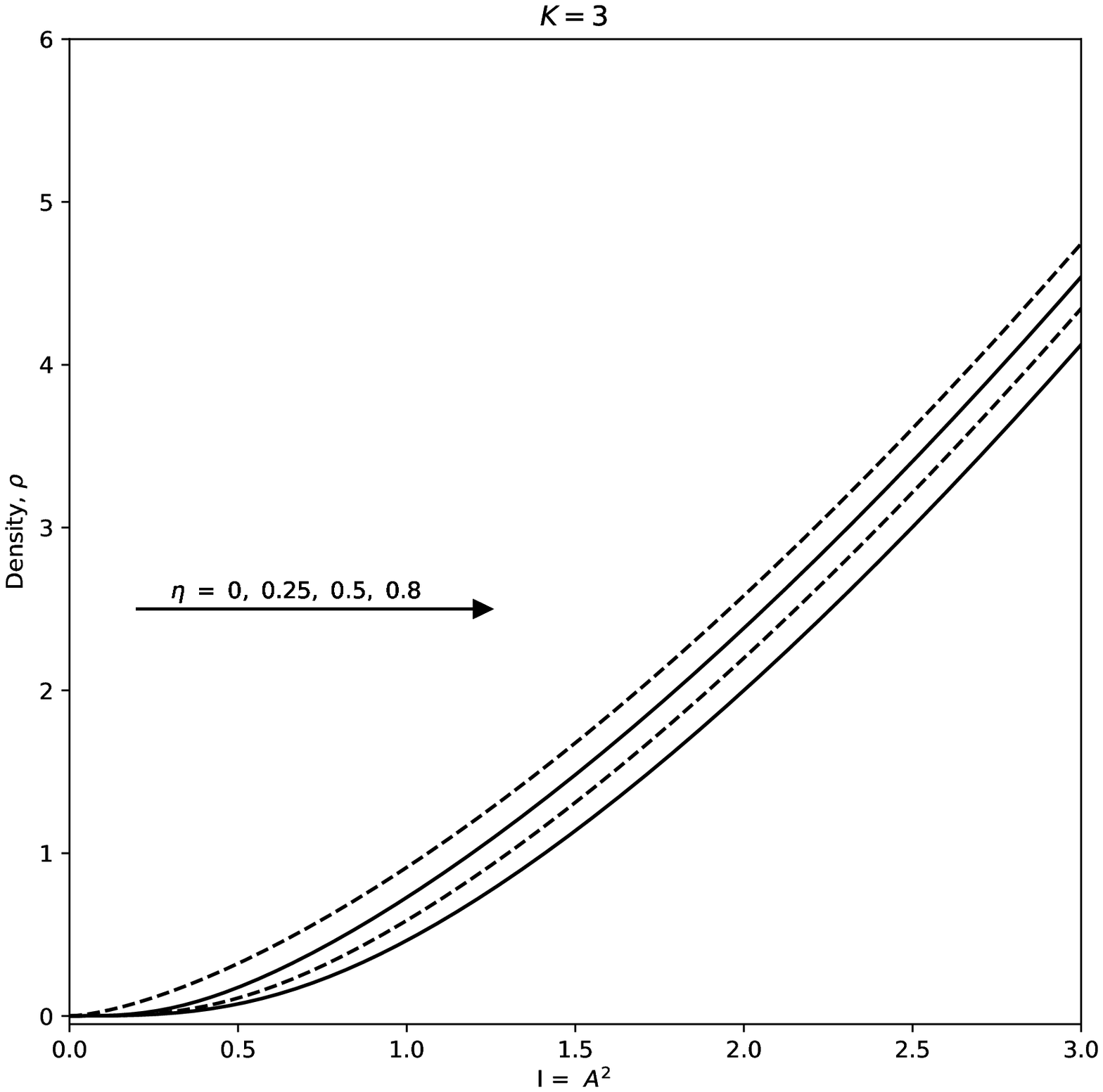}
\end{minipage}\\
\begin{minipage}{0.48\textwidth}
\includegraphics[width=3.in,height=2.in]{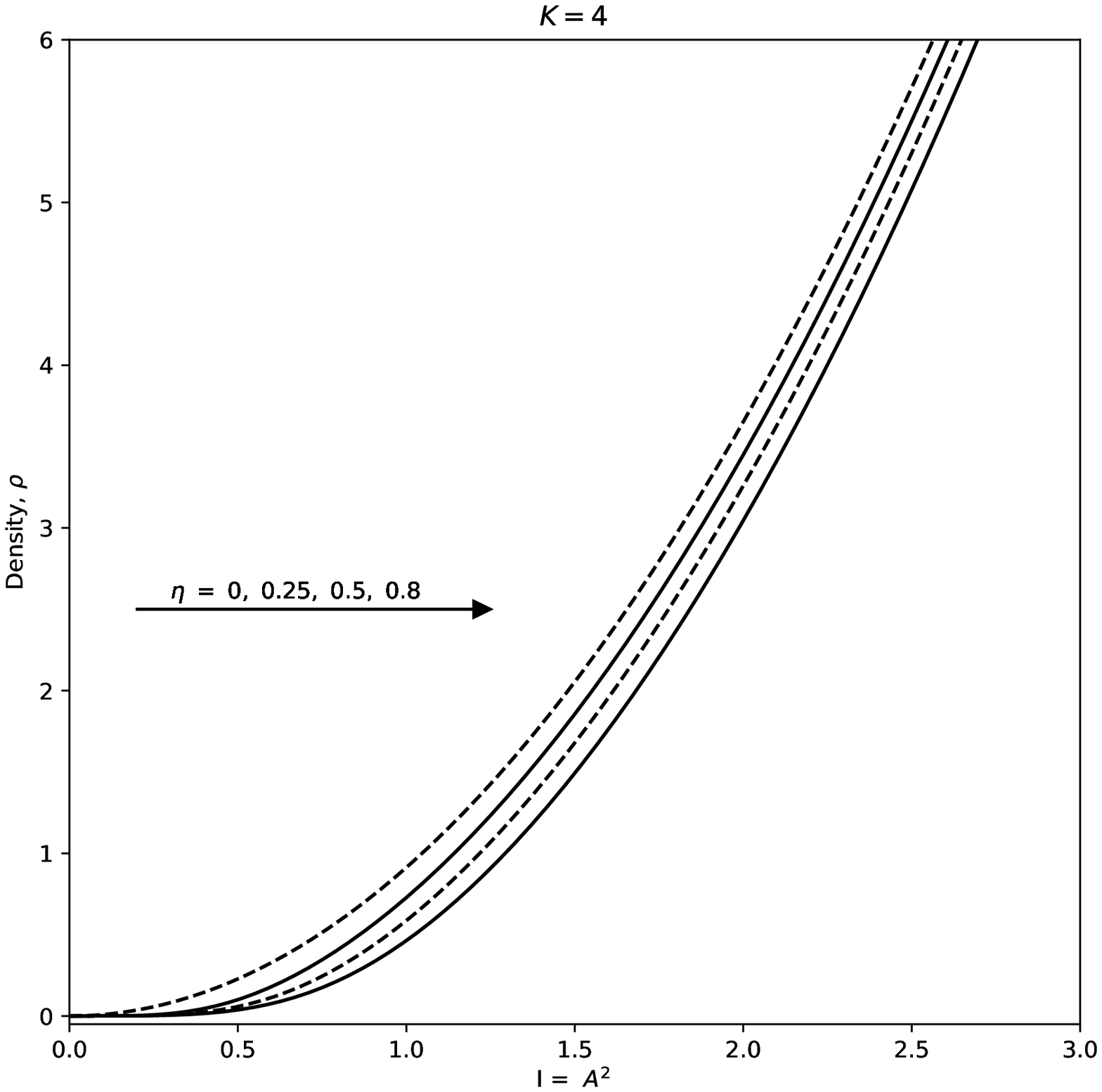}
\end{minipage}%
\begin{minipage}{0.48\textwidth}
\includegraphics[width=3.in,height=2.in]{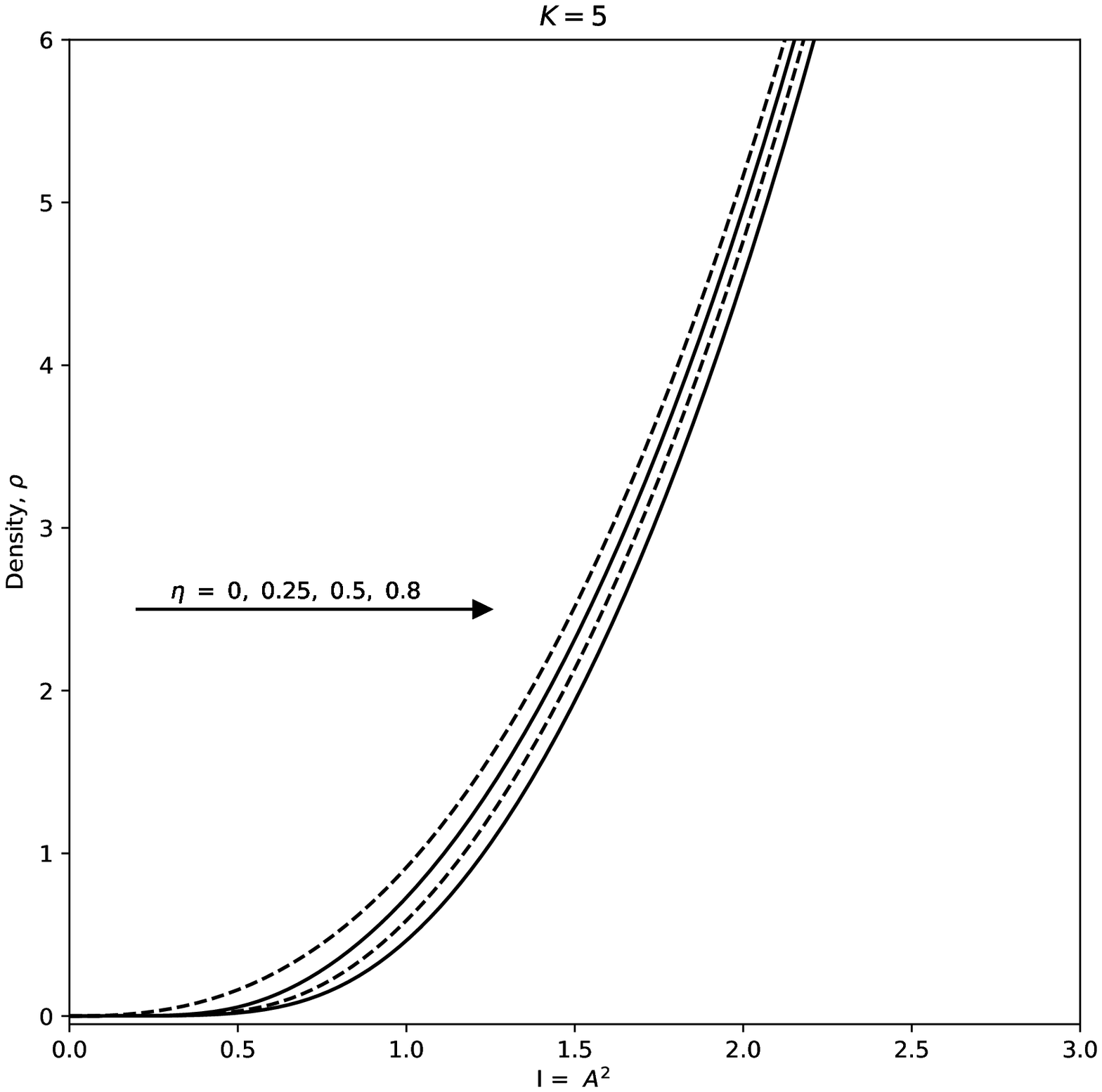}
\end{minipage}
\caption{\label{fig1}Variation of the fixed point of the plasma density $\rho$, given by eq. (\ref{fp2a}), as a function of the input power $I=A^2$ for $\eta=0$, $0.25$, $0.5$, $0.8$, and for four different values of the photon number $K$. We have used $\delta=-0.5$, $\sigma=1.5$, $\omega_0\tau_0=0.2$, $\mu=-0.25$. } 
 \end{figure}

 In the steady-state regime, solutions to the system of first-order ordinary differential equations (\ref{frs1})-(\ref{frs3}) are fixed points of the system. These solutions therefore correspond to the roots of the following set of three coupled multivariate polynomials in $A$, $M$ and $\rho$:
\begin{eqnarray}
0 &=&\frac{\gamma\rho}{\delta}  + \frac{\mu A^{2(K-1)}}{\delta} - \frac{2My}{A}, \label{frs11}\\
0 &=& \frac{\omega A}{\delta} + \frac{\gamma\omega_o\tau_o\rho A}{\delta} + \frac{\delta A M^2}{\delta} + \frac{\sigma A^3}{\delta}, \label{frs21}\\
0 &=& \nu A^2 \rho + \alpha A^{2K} -\eta \rho - a\rho^2, \label{frs31}
\end{eqnarray}
with in addition $y=0$ suggesting that at steady state, any arbitrary real constant will be a suitable solution for the laser amplitude. From eqs. (\ref{frs11})-(\ref{frs31}), we find the fixed-point solutions in $M$ and $\rho$ as being:
\begin{eqnarray}
M^2 &=&\frac{\omega_o\tau_o\mu A^{2(K-1)}-\sigma  A^2-\omega }{\delta}. \label{fp1}\\
\rho &=& \frac{-\eta \pm \sqrt{\eta^2 + 4a\bigg[\alpha-\frac{\nu\mu}{\gamma}\bigg]A^{2K}}}{2a}. \label{fp2}
\end{eqnarray}
Instructively formula (\ref{fp2}) suggests two possible fixed-point solutions for the plasma density, i.e. a negative-valued solution and a positive solution. Since $\rho$ must be positive or zero to be physically significant, the negative solution is physically irrelevant. We must therefore keep the solution with positive sign, such that the fixed-point solution for the plasma density is: 
\begin{equation}
\rho = \frac{-\eta + \sqrt{\eta^2 + 4a\bigg[\alpha-\frac{\nu\mu}{\gamma}\bigg]A^{2K}}}{2a}. \label{fp2a}
\end{equation}
The fixed point of $\rho$ will be zero when $\nu$ (the coefficient of avalanche impact ionization), $\alpha$ (the coefficient of plasma balance due to multiphoton absorption), $\mu$ (the mutiphoton ionization coefficient) and $\gamma$ (the strength of coupling of the electron plasma to the optical field), are related through:
\begin{equation}
 \alpha=\frac{\nu\mu}{\gamma}, \label{cond1}
\end{equation}
for nonzero values of the electron-hole radiative recombination coefficient $a$ and the electron diffusion coefficient $\eta$. Note that the steady-state value of $\rho$ given by eq. (\ref{fp2a}), implies that $\alpha$ should be greater than $\nu\mu/\gamma$ and the electron-hole radiative recombination coefficient $a$ should be positive. Since the multiphoton ionization coefficient $\mu$ can be negtive, choosing $\alpha$, $\nu$ and $\gamma$ all positive will be enough to ensure a physically consistent value of $\rho$ at steady state. Fig. \ref{fig1} shows the variation of the fixed point eq. (\ref{fp2a}), as a function of the input power $I=A^2$ of femtosecond laser for four different values of $K$, and some values of the electron diffusion coefficient $\eta$. Values of other characteristic parameters of the model are indicated in the caption.
According to fig. \ref{fig1} the fixed point of the plasma density increases when the input power $A^2$ is increased, irrespective of the value of $K$. However, an increase of $K$ affects the law of dependence of $\rho$ with the input power. Indeed, fig. \ref{fig1} shows that $\rho$ is more and more strongly nonlinearly dependent on $A^2$. \par The increase of the fixed-point value of the plasma density with increase of the input power, can be understood by invoking the fact that the higher the intensity of femtosecond laser, the higher the amount of energy stored by the femtosecond laser in the material. Considering such higher energy stored in the material, photoionization processes should be stronger and consequently the density of the electron plasma too should be higher. When both $K$ and $\eta$ are very small, the fixed-point solution $\rho$ is seen to increase linearly with the laser input power $A^2$. However, as $K$ is increased the fixed-point value of $\rho$ turns gradually from linear to parabolic (nonlinear) in $A^2$. The flexion marking the deviation to a parabolic curve is, as evidenced in figure \ref{fig1}, more and more sharp as the electron diffusion coefficient $\eta$ is increased. It is to be remarked that in the nonlinear regime, the fixed point of the plasma density at the same input laser intensity will be lowered down as $\eta$ increases. These behaviours clearly suggest that the electron diffusion processes play a key role in the process, namely the fixed-point analysis carried out in this section suggests that the relationship between the threshold value of the plasma density and the input laser intensity can be controlled by electron diffusion processes.  

\section{Soliton-train solutions}
\label{four}
In section \ref{three}, fixed-point solutions to the coupled first-order ordinary differential equations (\ref{frs1})-(\ref{frs4}), describing the laser dynamics together with temporal evolution of the plasma density, were examined. We obtained that any arbitrary real constant could be a fixed point of the laser amplitude, whereas the fixed-point solution for the plasma density $\rho$ was a function of the input laser intensity $A^2$. The dependence of $\rho$ on $A^2$ emerged to be linear for small $K$ and $\eta$, however increasing $K$ and $\eta$ simultaneously, turns the linear dependence into a parabolic function reflecting a nonlinear variation of the steady-state value of the plasma density with the input power. \par
\begin{figure}\centering
\begin{minipage}{0.5\textwidth}
\includegraphics[width=3.in,height=2.in]{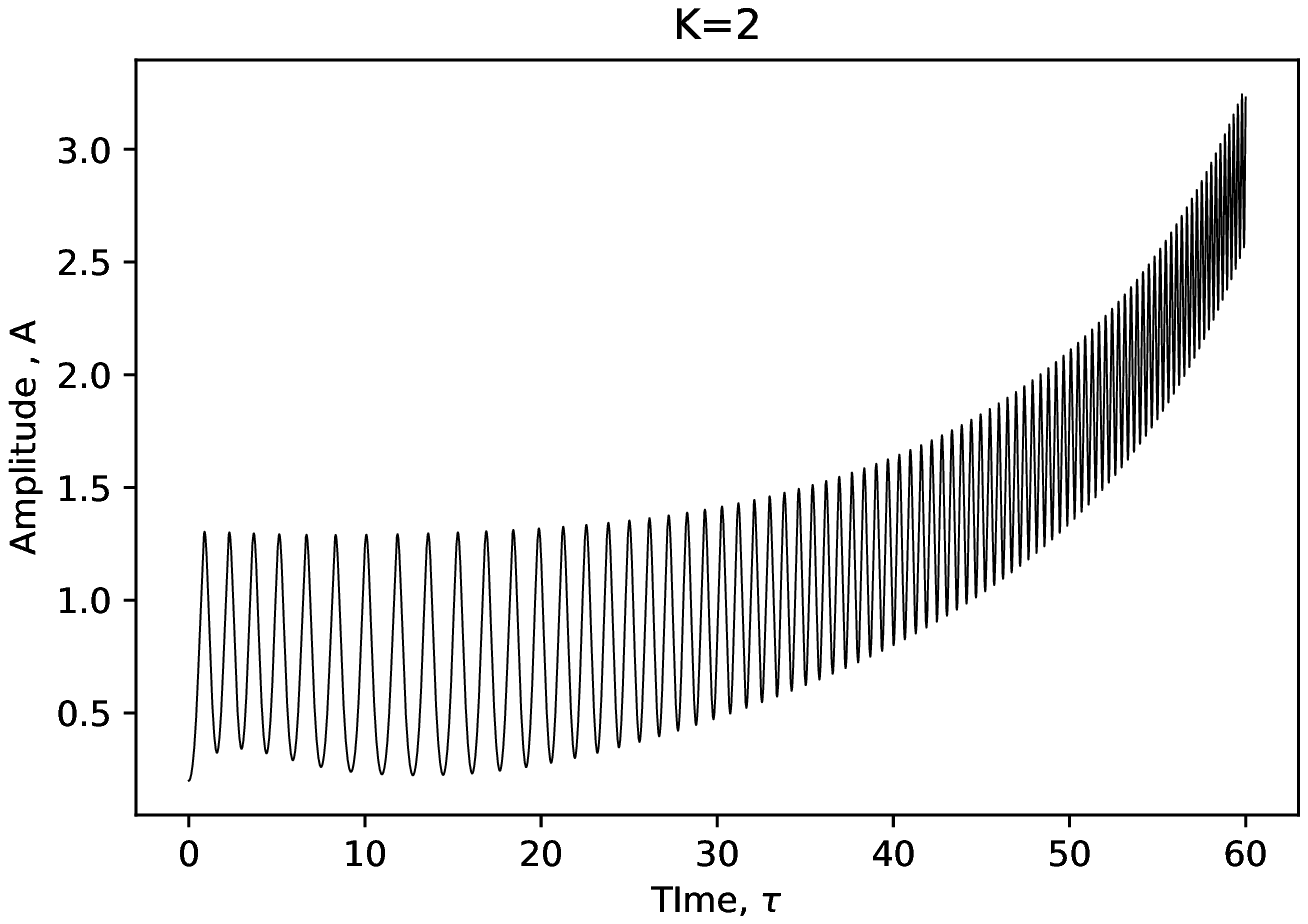} 
\end{minipage}%
\begin{minipage}{0.5\textwidth}
\includegraphics[width=3.in,height=2.in]{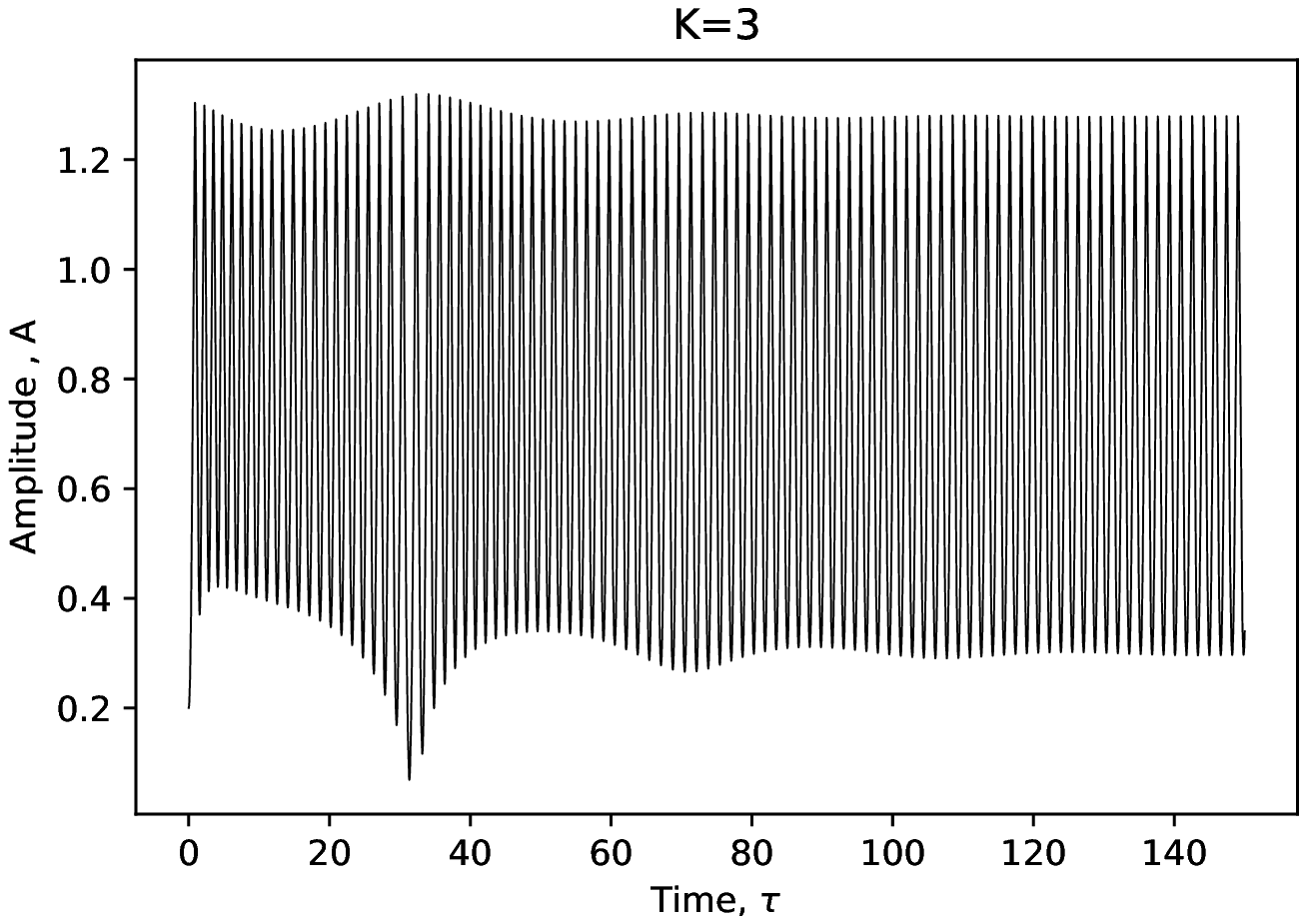} 
\end{minipage}\\
\begin{minipage}{0.5\textwidth}
\includegraphics[width=3.in,height=2.in]{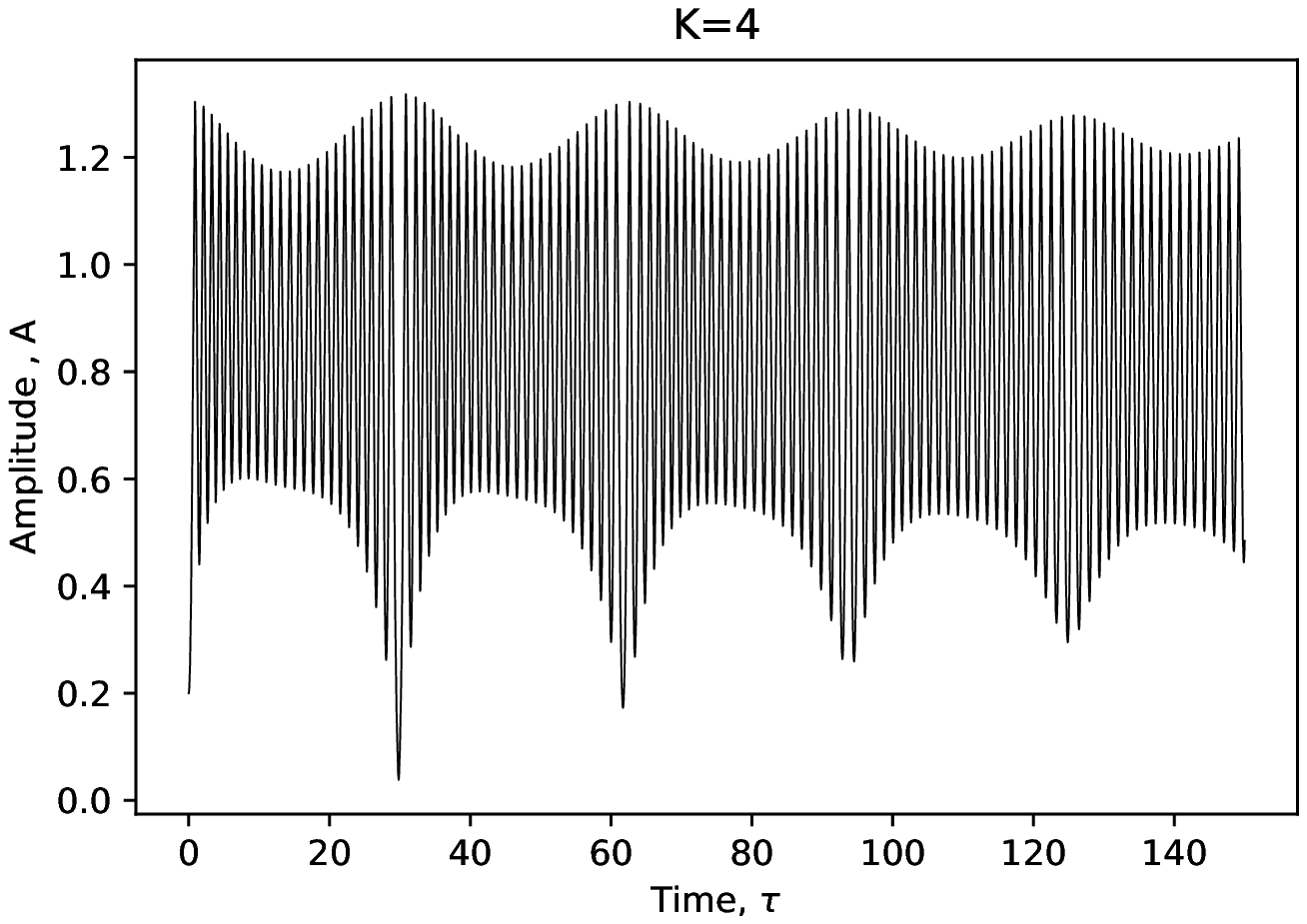} 
\end{minipage}%
\begin{minipage}{0.5\textwidth}
\includegraphics[width=3.in,height=2.in]{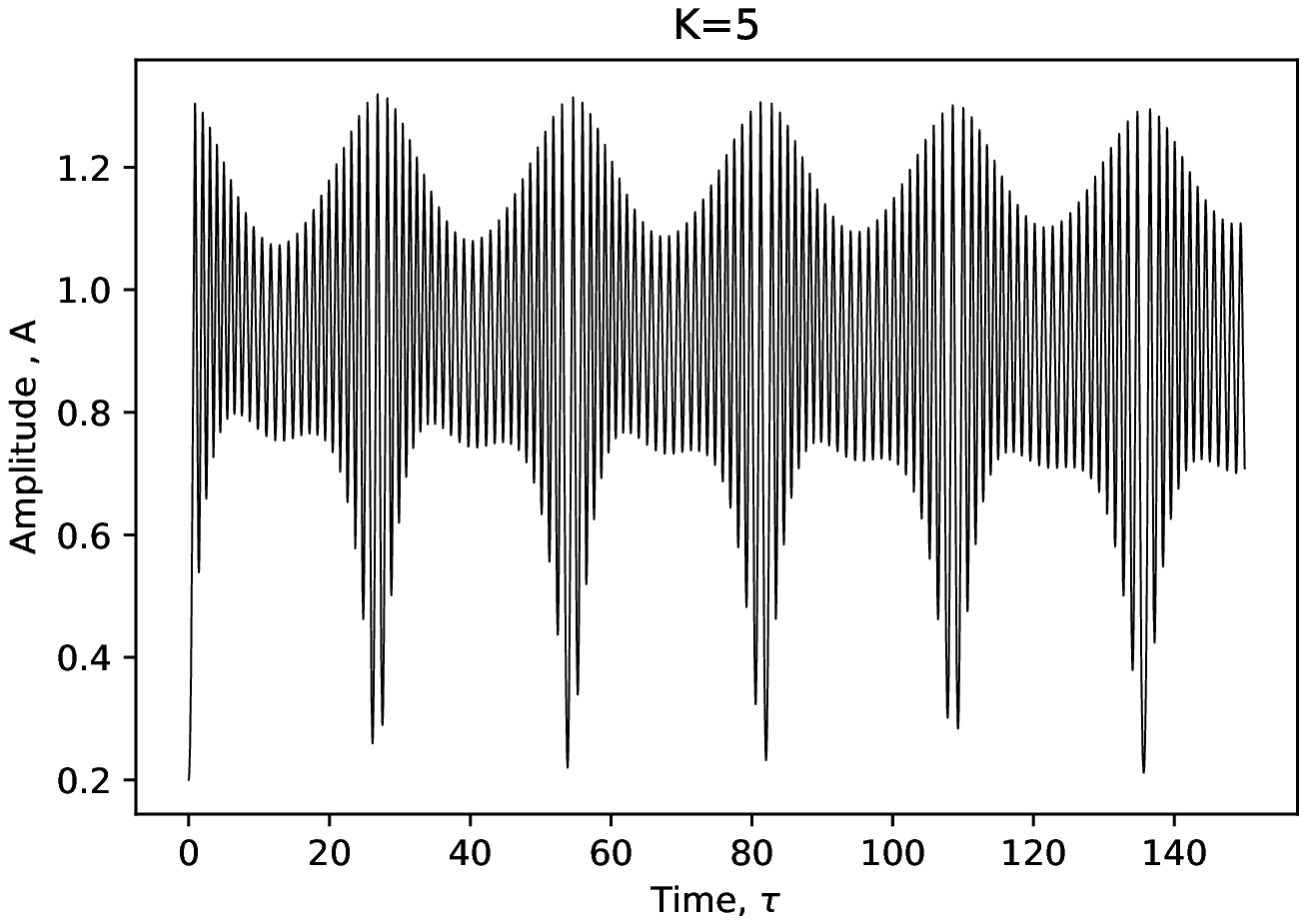}
\end{minipage}
\caption{\label{fig2} Temporal profile of the laser amplitude $A(\tau)$, generated numerically for different values of $K$ and for $\eta= 0.005$.} 
\end{figure}

\begin{figure}\centering
\begin{minipage}{0.5\textwidth}
\includegraphics[width=3.in,height=2.in]{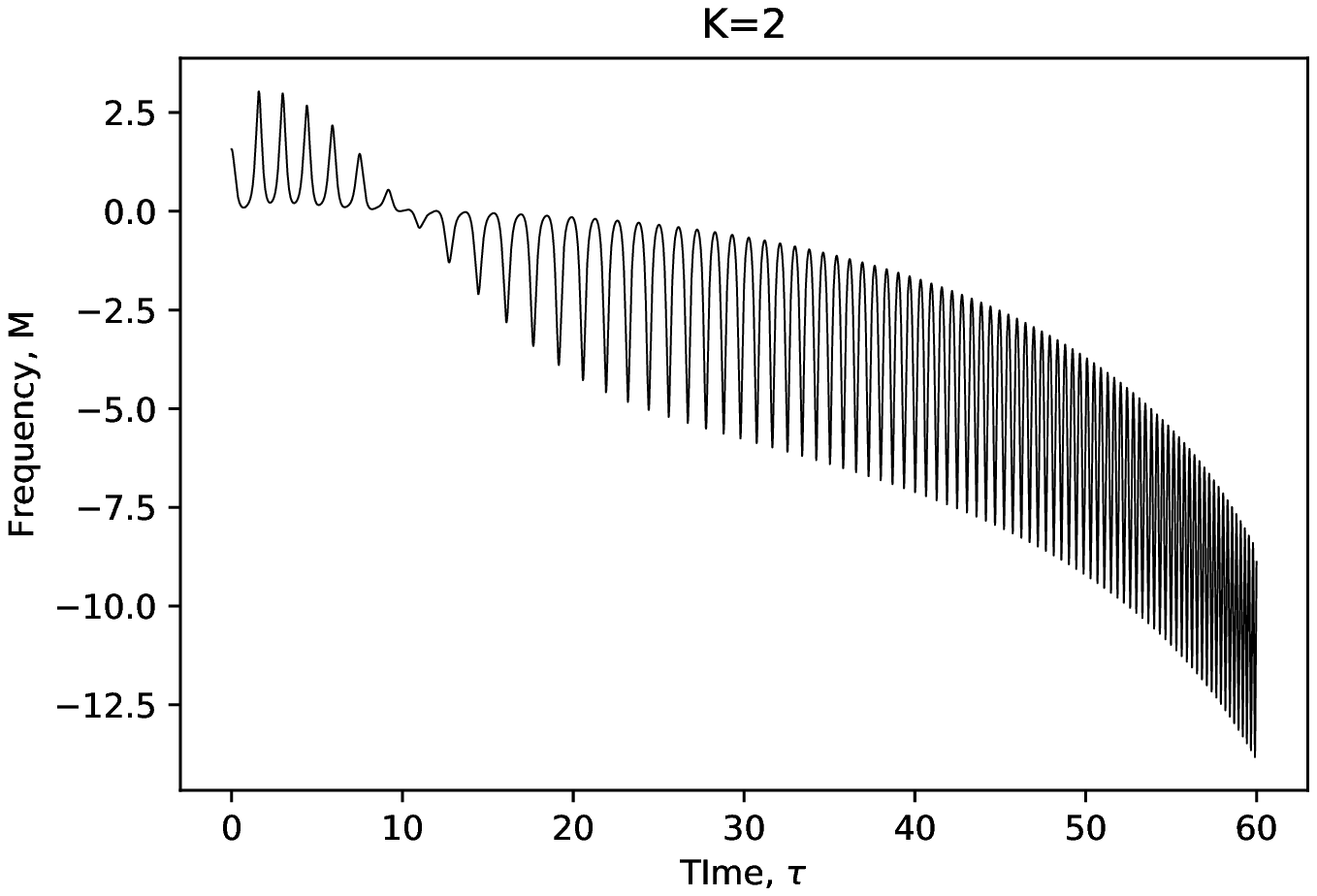} 
\end{minipage}%
\begin{minipage}{0.5\textwidth}
\includegraphics[width=3.in,height=2.in]{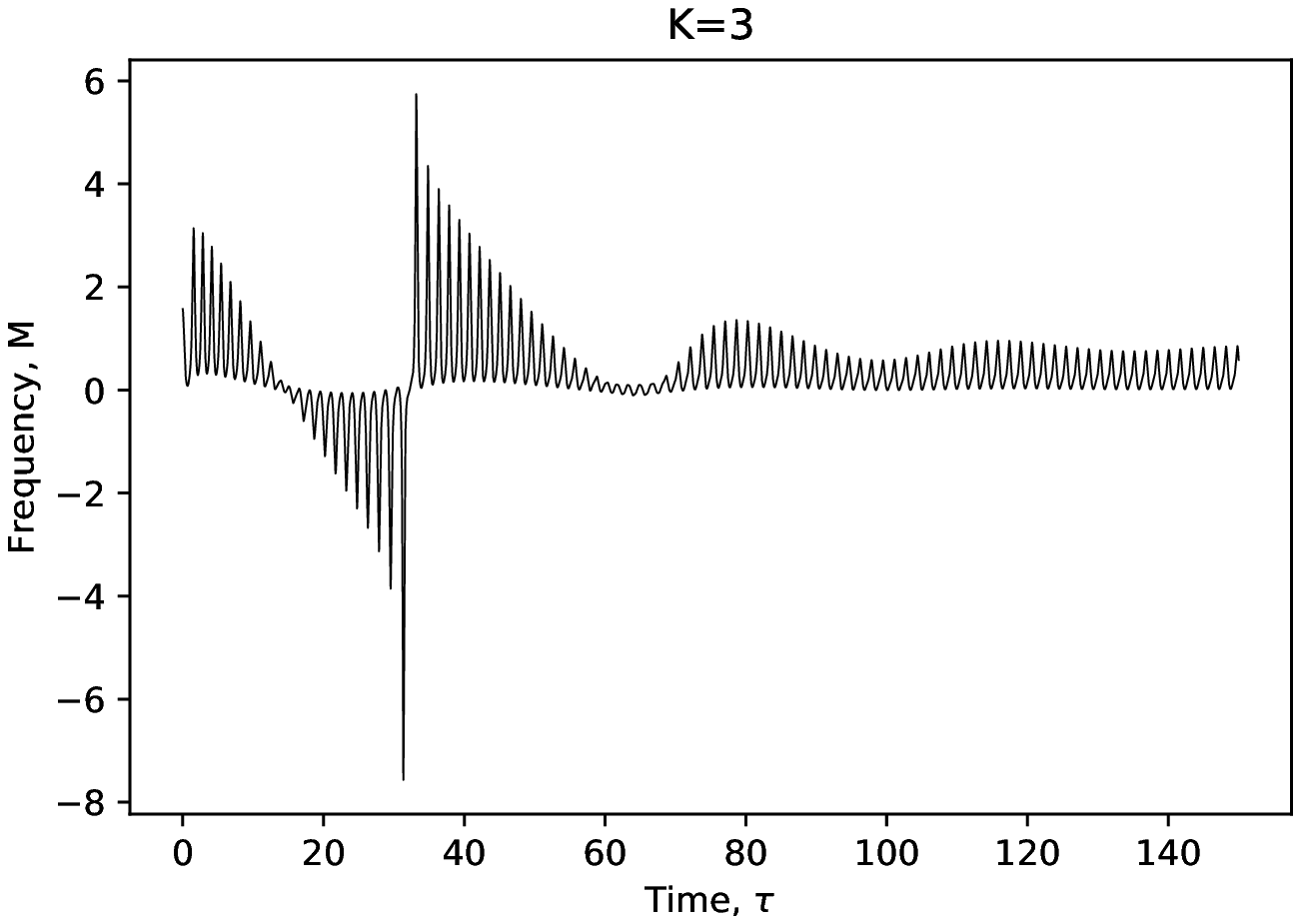} 
\end{minipage}\\
\begin{minipage}{0.5\textwidth}
\includegraphics[width=3.in,height=2.in]{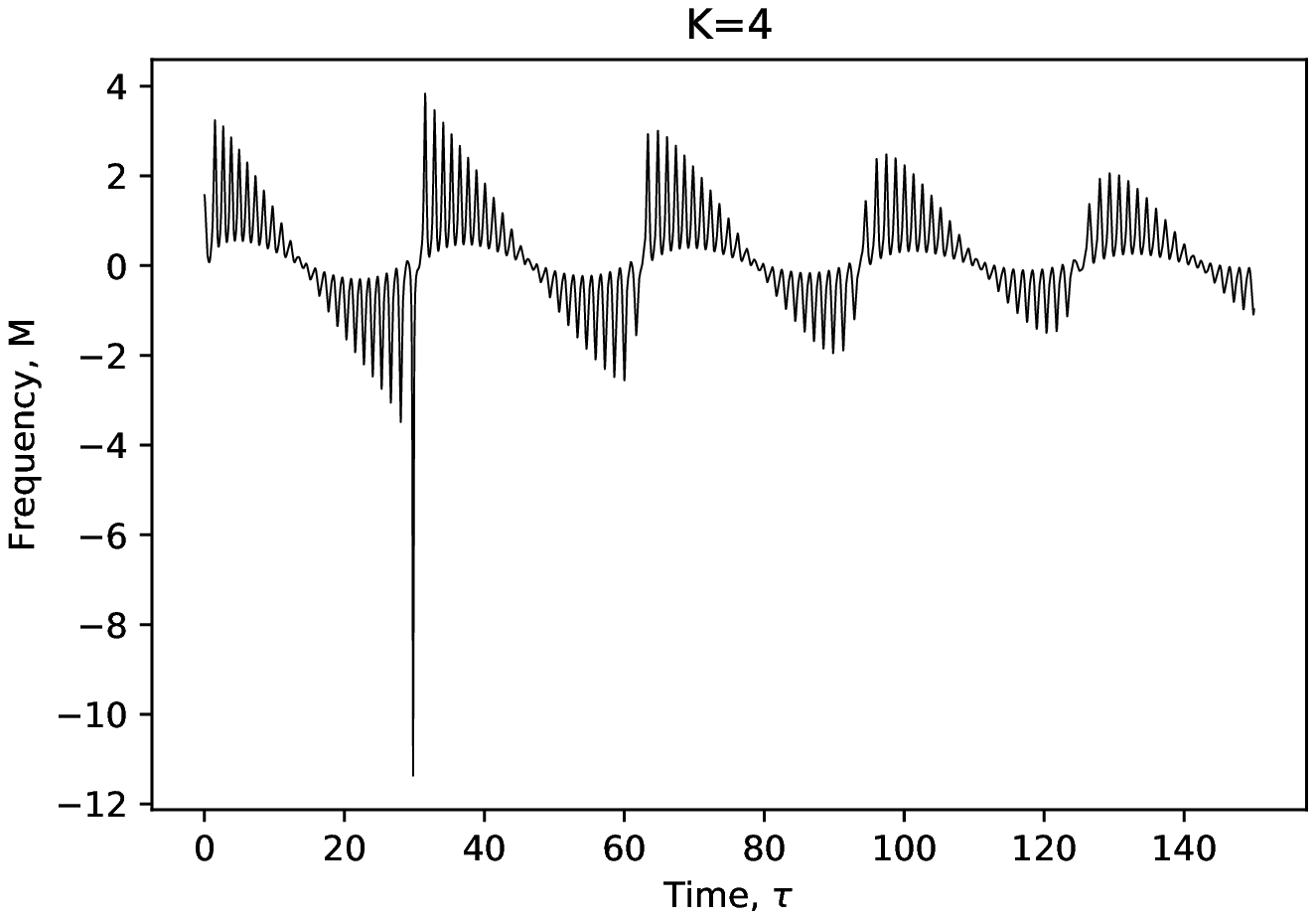} 
\end{minipage}%
\begin{minipage}{0.5\textwidth}
\includegraphics[width=3.in,height=2.in]{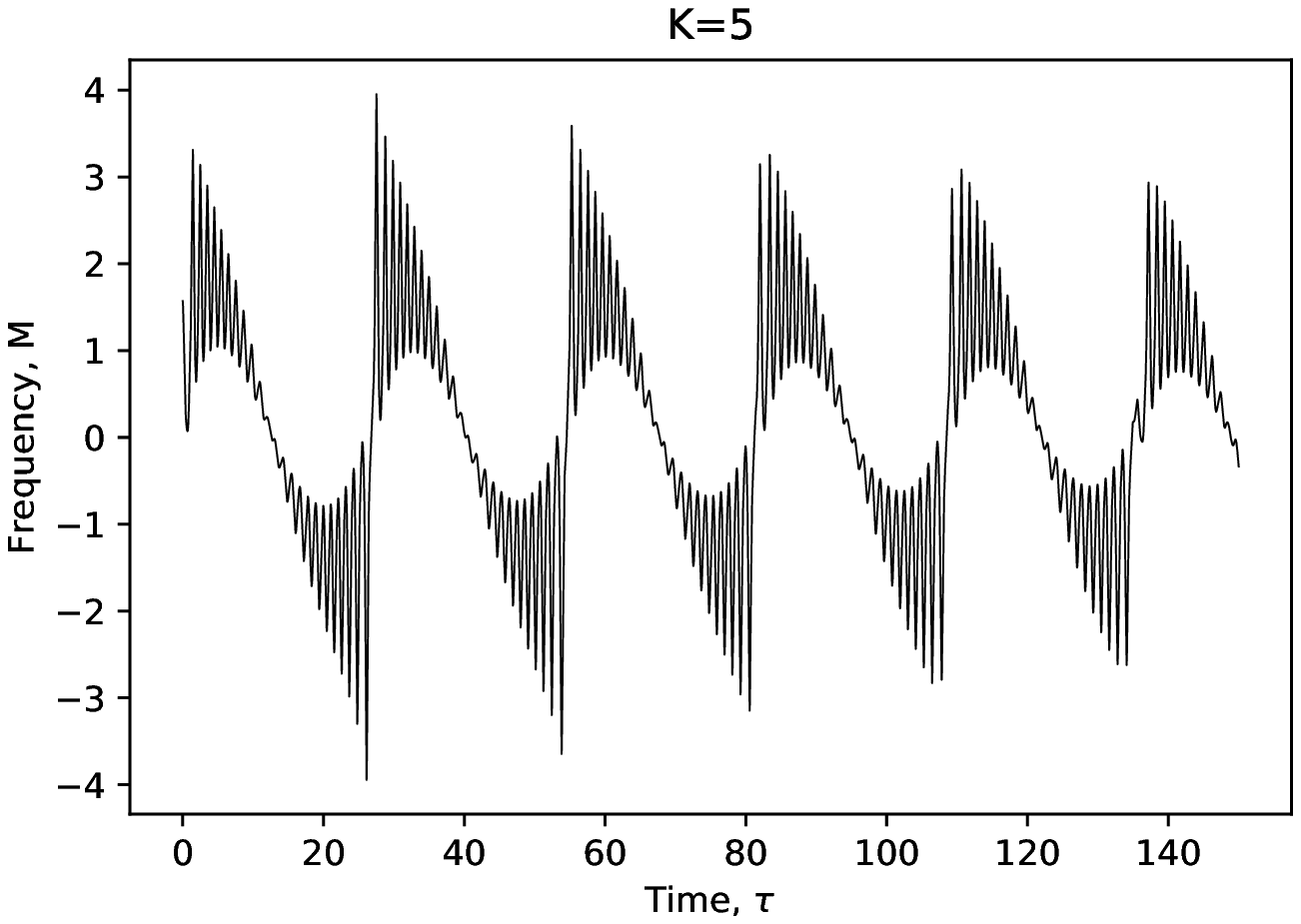}
\end{minipage}
\caption{\label{fig3}Temporal profile of the instantaneous frequency $M(\tau)$ of the laser, generated numerically for different values of $K$ and for $\eta=0.005$.} 
\end{figure}

\begin{figure}\centering
\begin{minipage}{0.5\textwidth}
\includegraphics[width=3.in,height=2.in]{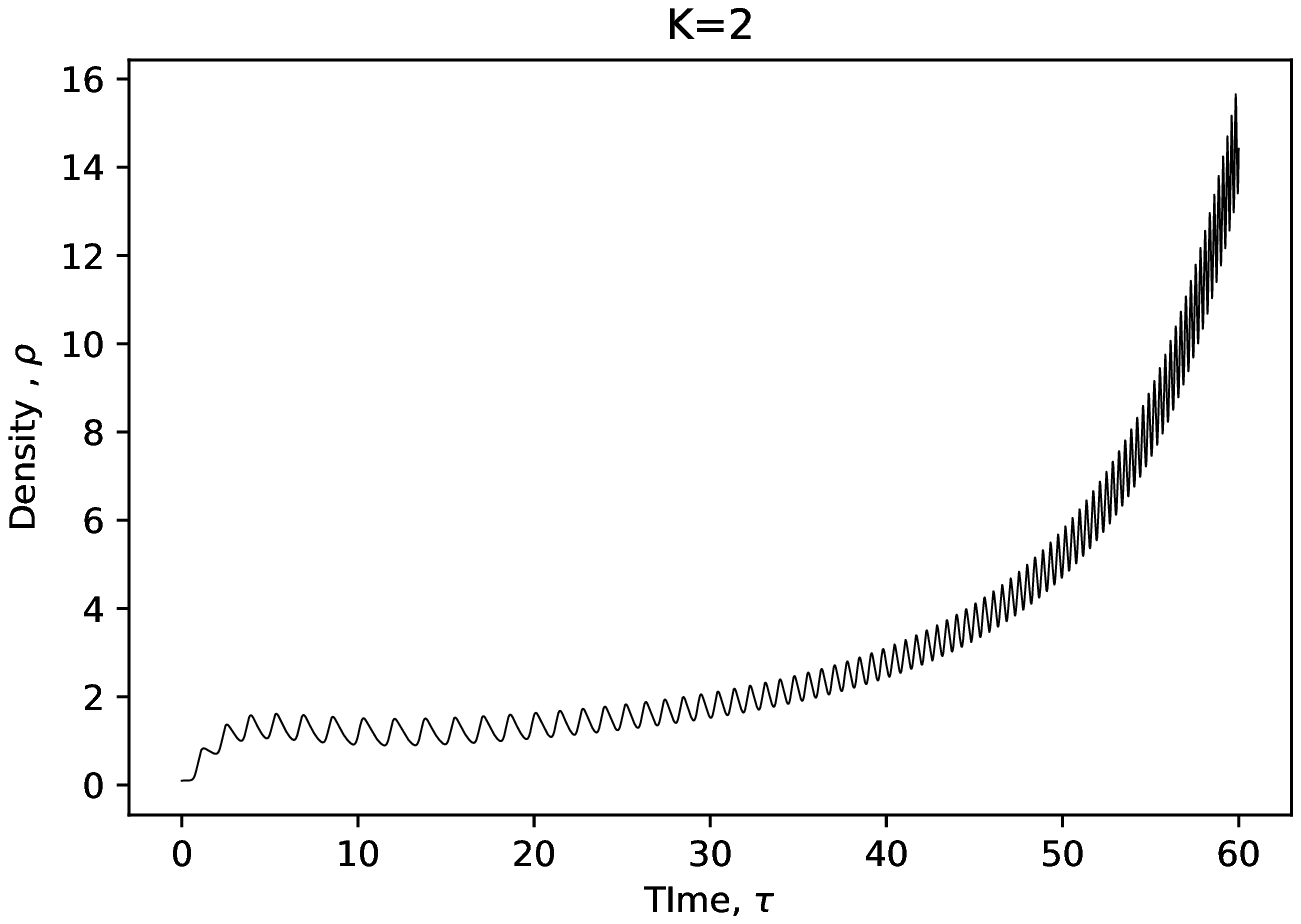} 
\end{minipage}%
\begin{minipage}{0.5\textwidth}
\includegraphics[width=3.in,height=2.in]{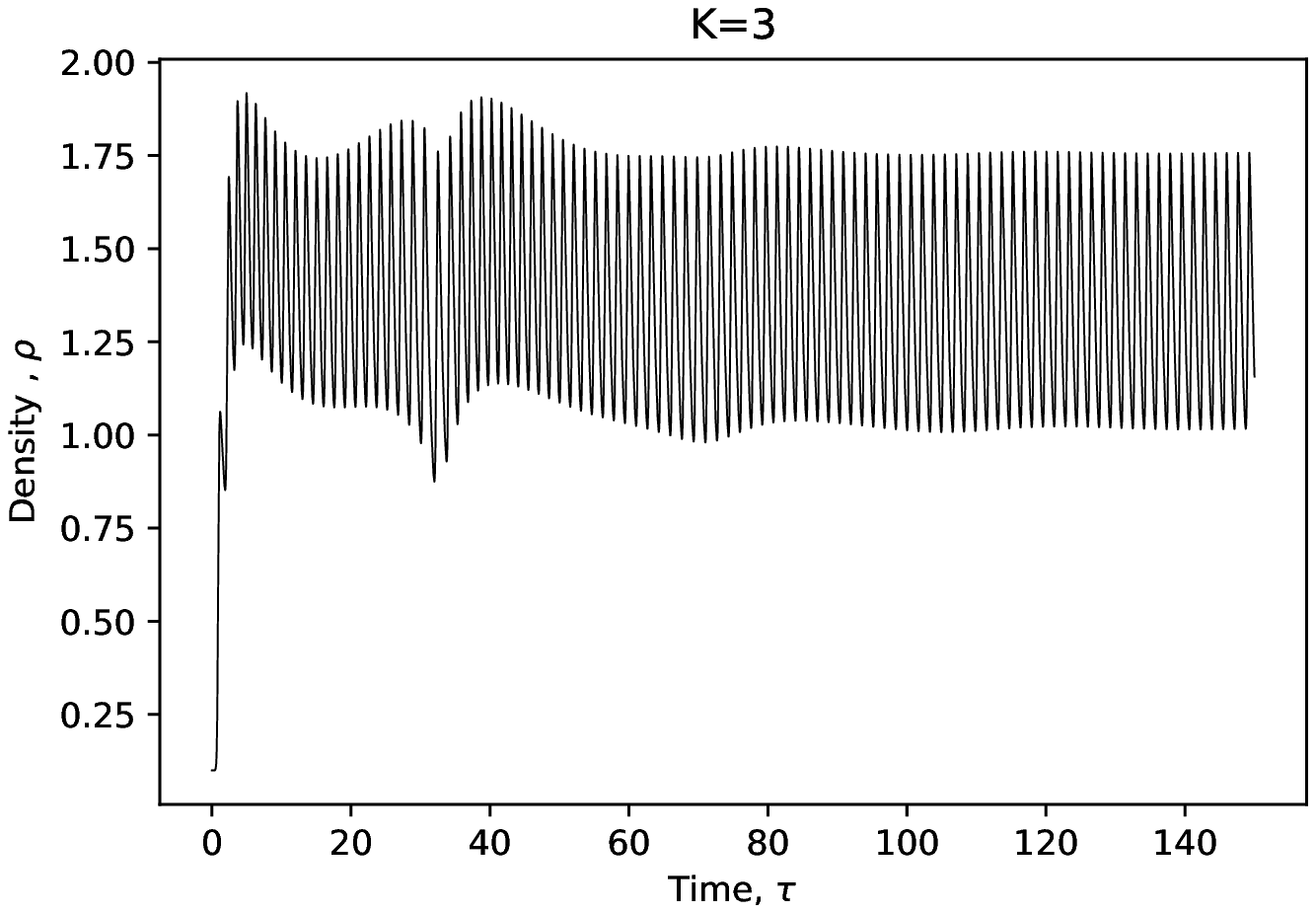} 
\end{minipage}\\
\begin{minipage}{0.5\textwidth}
\includegraphics[width=3.in,height=2.in]{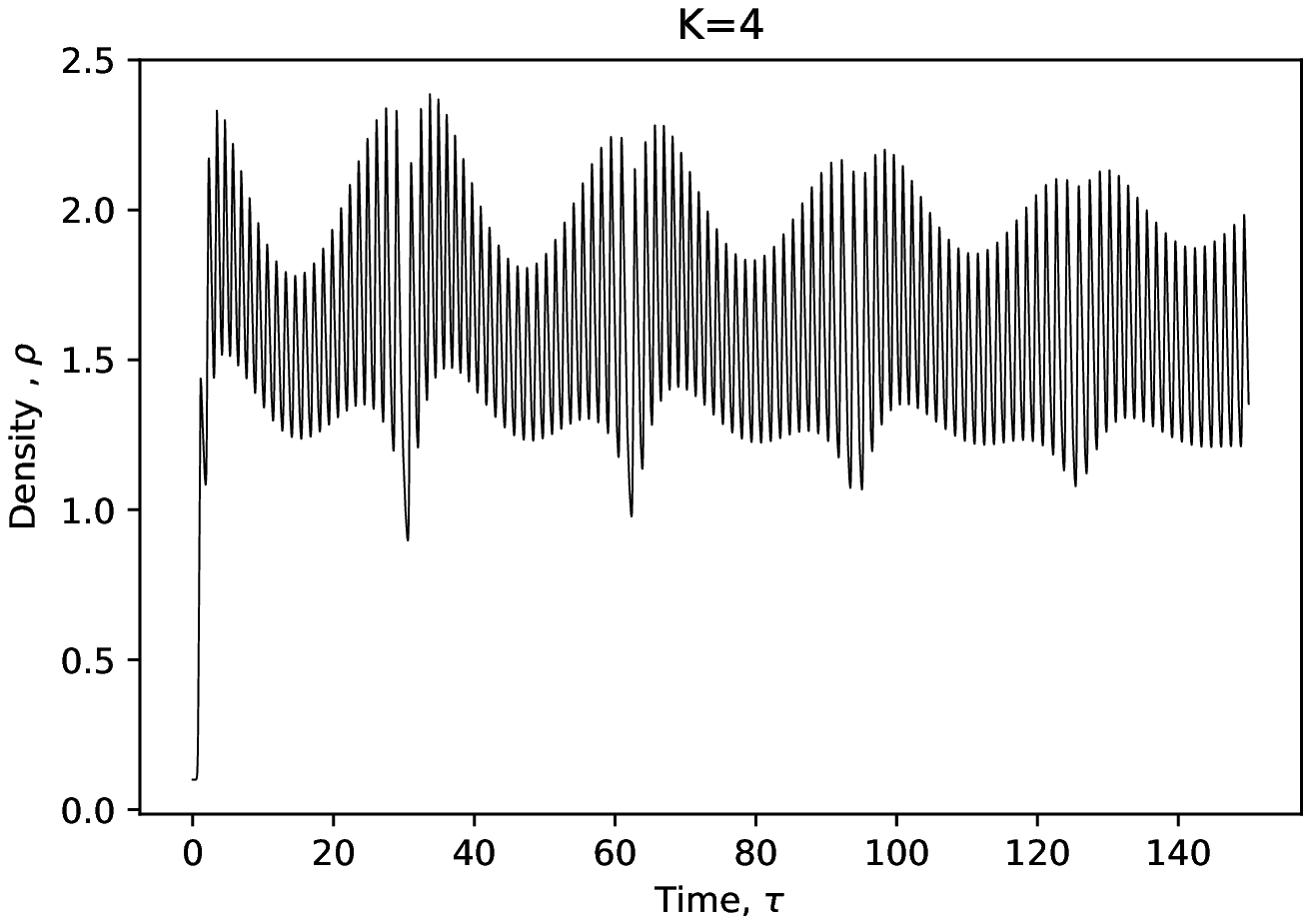} 
\end{minipage}%
\begin{minipage}{0.5\textwidth}
\includegraphics[width=3.in,height=2.in]{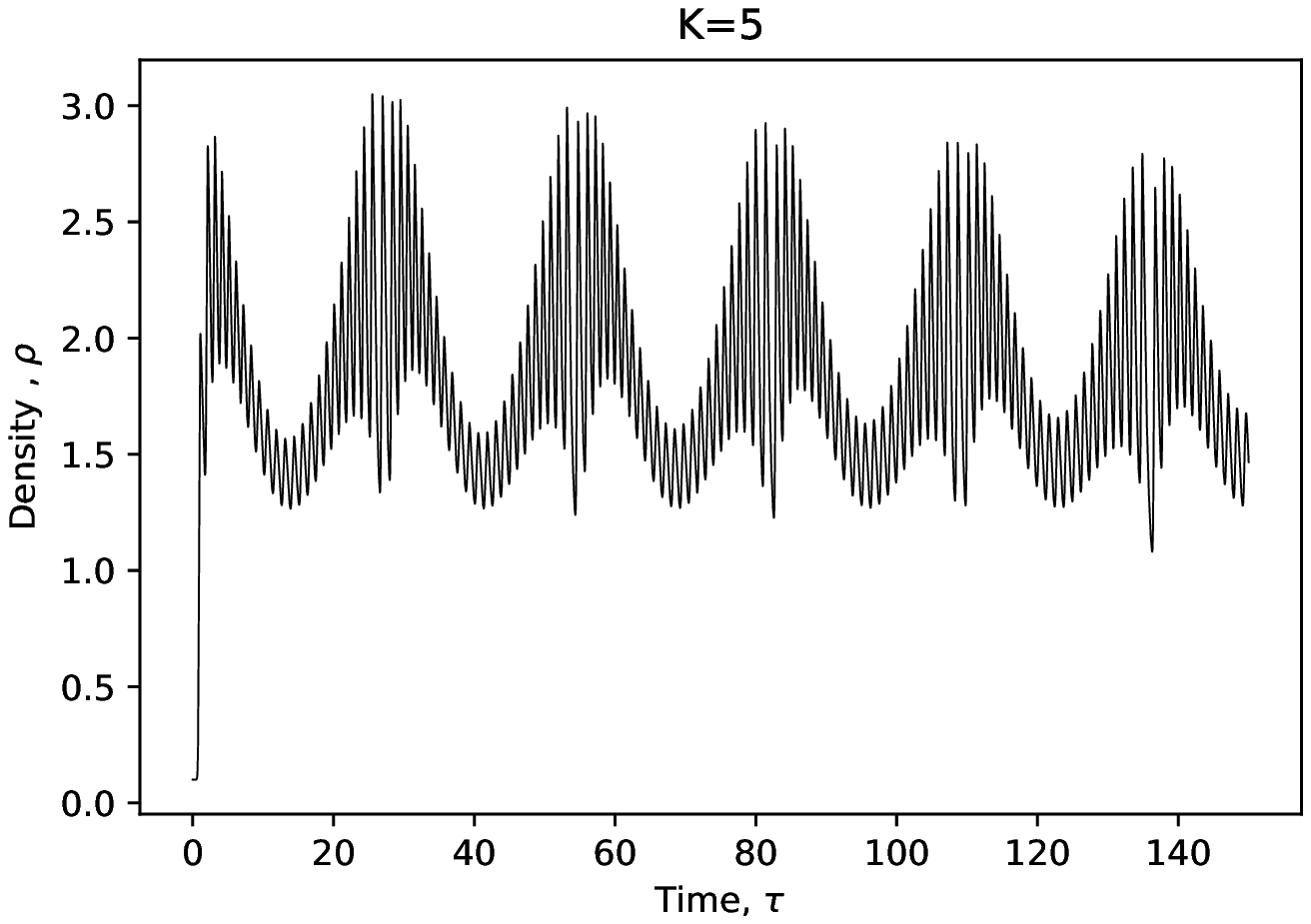}
\end{minipage}
\caption{\label{fig4} Temporal profile of the plasma density $\rho(\tau)$, generated numerically for different values of $K$ and for $\eta= 0.005$.} 
\end{figure}

 
In this section, we wish to look at temporal profiles of the laser amplitude $A(\tau)$ and of the plasma density $\rho(\tau)$. For this purpose, we solved numerically the set of coupled first-order ordinary differential equations (\ref{frs1})-(\ref{frs4}), using an explicit sixth-order Runge-Kutta scheme \cite{luth}. To highlight the influence of electron diffusion processes on shape profiles of the laser amplitude and the plasma density, the numerical solutions $A(\tau)$ and $\rho(\tau)$ were generated considering a very small value of $\eta$ and a relative large value of this parameter. In figs. (\ref{fig2}), (\ref{fig3}) and (\ref{fig4}), we display time series of $A(\tau)$, $M(\tau)$ and $\rho(\tau)$ corresponding to temporal profiles of the laser amplitude, the laser instantaneous frequency $M(\tau)=\partial \phi/\partial \tau$ and plasma density $\rho(\tau)$, for four different values of $K$ (i.e. $K=2$, 3, 4, 5) and for $\eta=0.005$. 
Fig. \ref{fig2} features a laser amplitude $A(\tau)$ forming a periodic train of anharmonic oscillations, with an exponential amplification as time goes when $K=2$. An increase of $K$ gradually softens the exponental amplification and for large values of $K$, the laser amplitude deploys as a regular train of quasi-periodic temporal pulses. Likewise, the instantaneous frequency $M$ and the plasma density $\rho$, shown respectively in figs. \ref{fig3} and \ref{fig4}, evolve from irregular to regular quasi-periodic nonlinear wave patterns as $K$ is increased from 2 to 5. Note the very small damping of the plasma density with time for $K=5$, contrasting with the very strong exponential amplification observed in the first graph of the same figure when $K=2$. 
\par
\begin{figure}\centering
\begin{minipage}{0.5\textwidth}
\includegraphics[width=3.in,height=2.in]{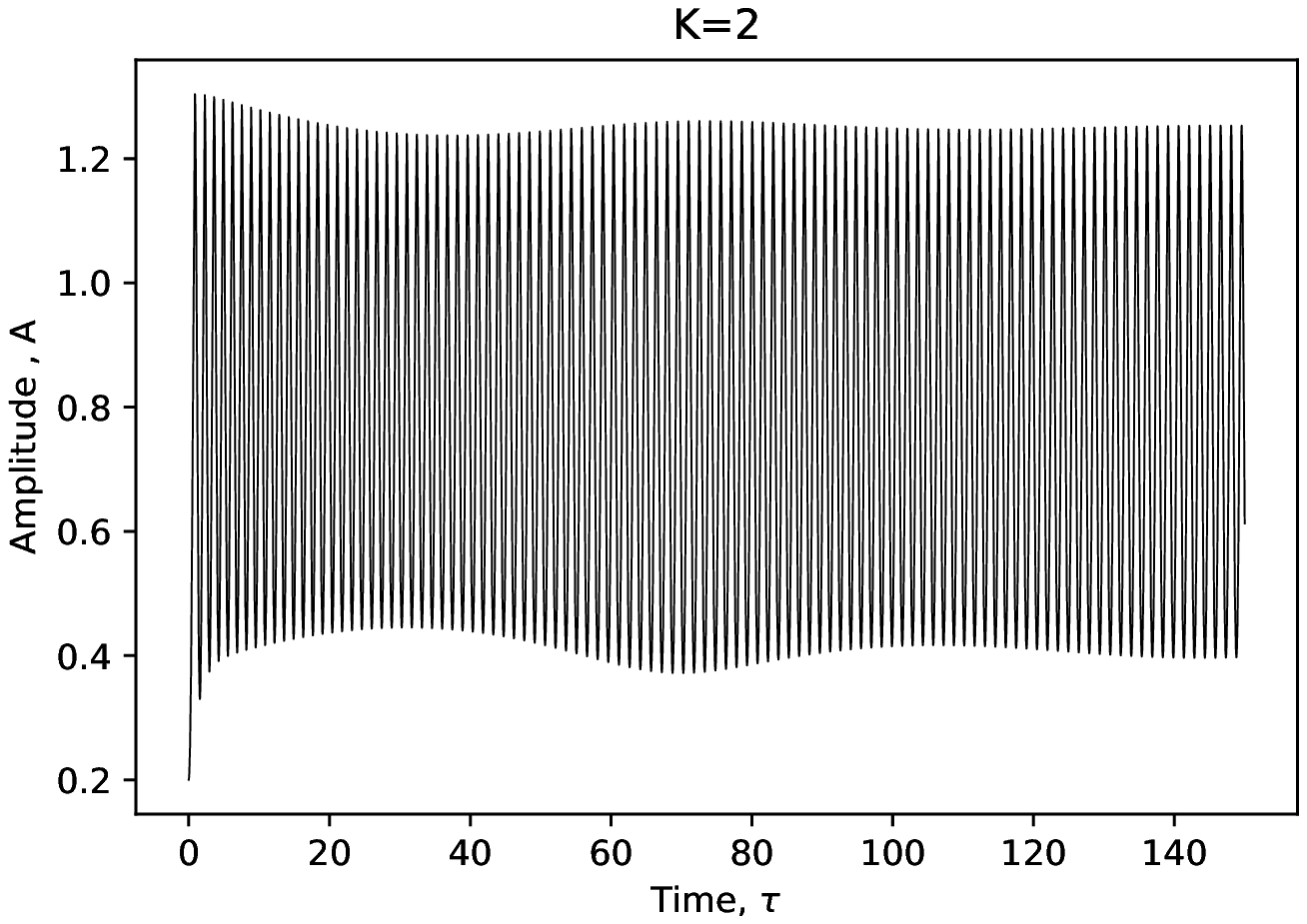} 
\end{minipage}%
\begin{minipage}{0.5\textwidth}
\includegraphics[width=3.in,height=2.in]{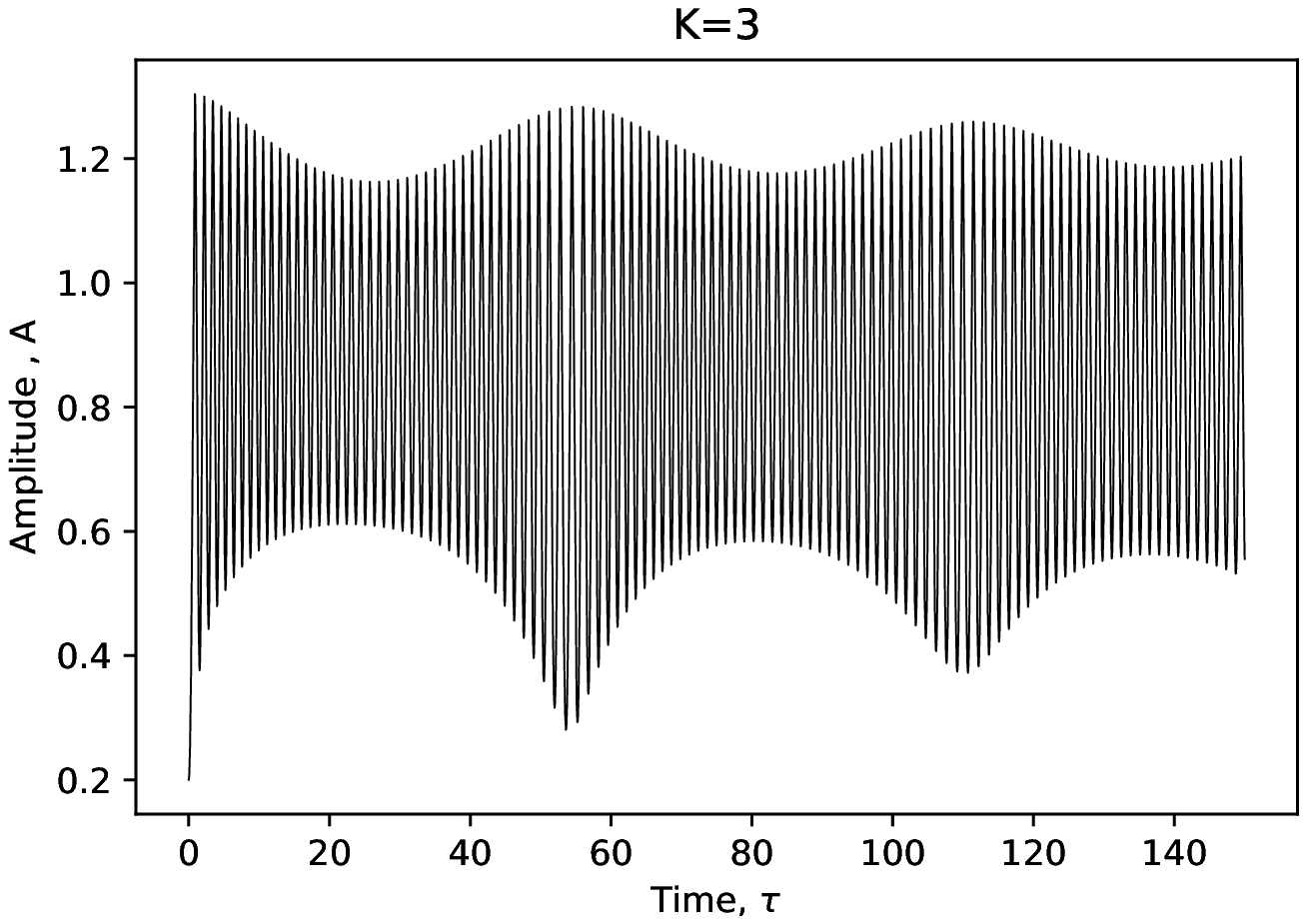} 
\end{minipage}\\
\begin{minipage}{0.5\textwidth}
\includegraphics[width=3.in,height=2.in]{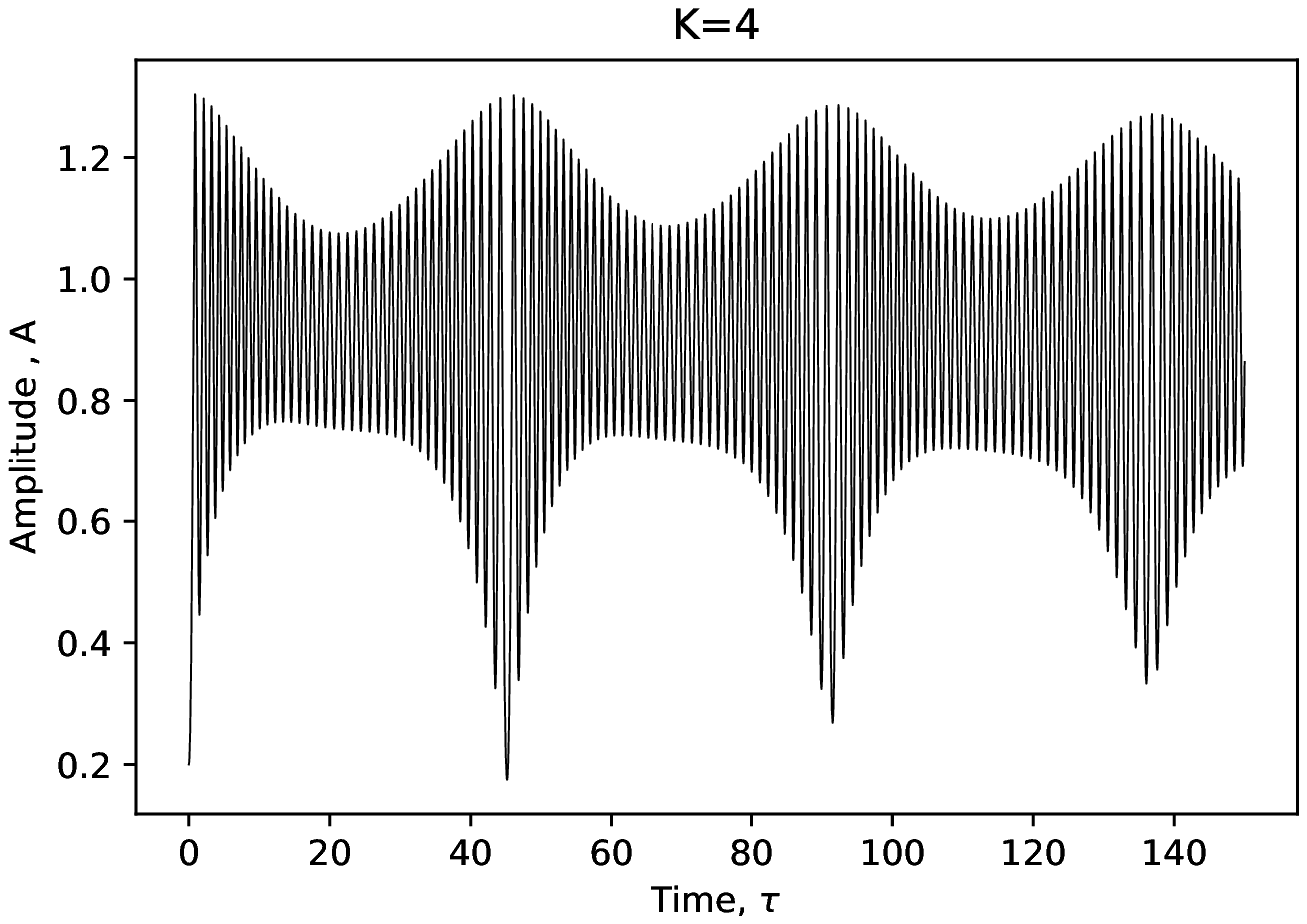} 
\end{minipage}%
\begin{minipage}{0.5\textwidth}
\includegraphics[width=3.in,height=2.in]{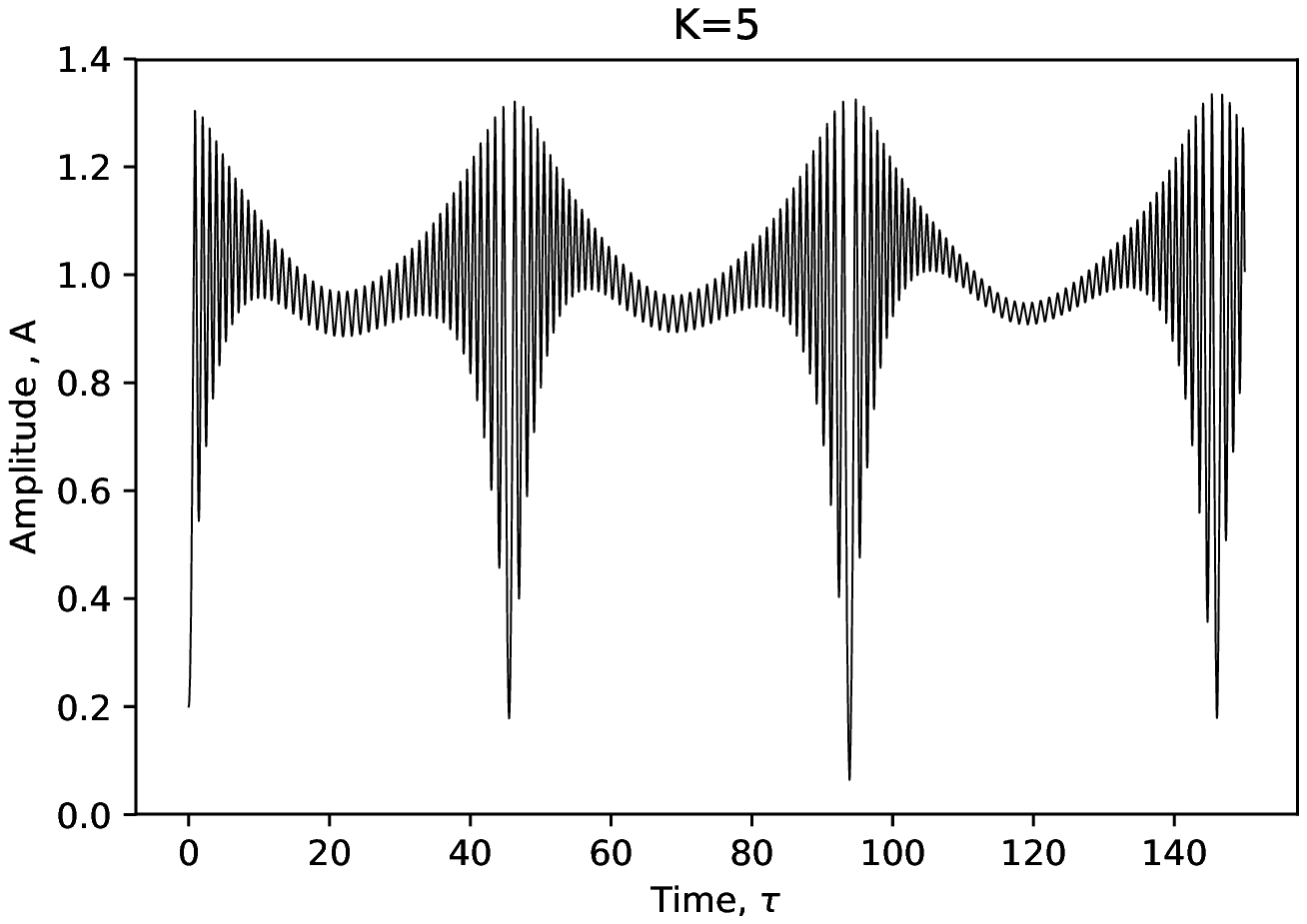}
\end{minipage}
\caption{\label{fig5} Temporal profile of the laser amplitude $A(\tau)$, generated numerically for different values of $K$ and for $\eta= 0.25$.} 
\end{figure}

\begin{figure}\centering
\begin{minipage}{0.5\textwidth}
\includegraphics[width=3.in,height=2.in]{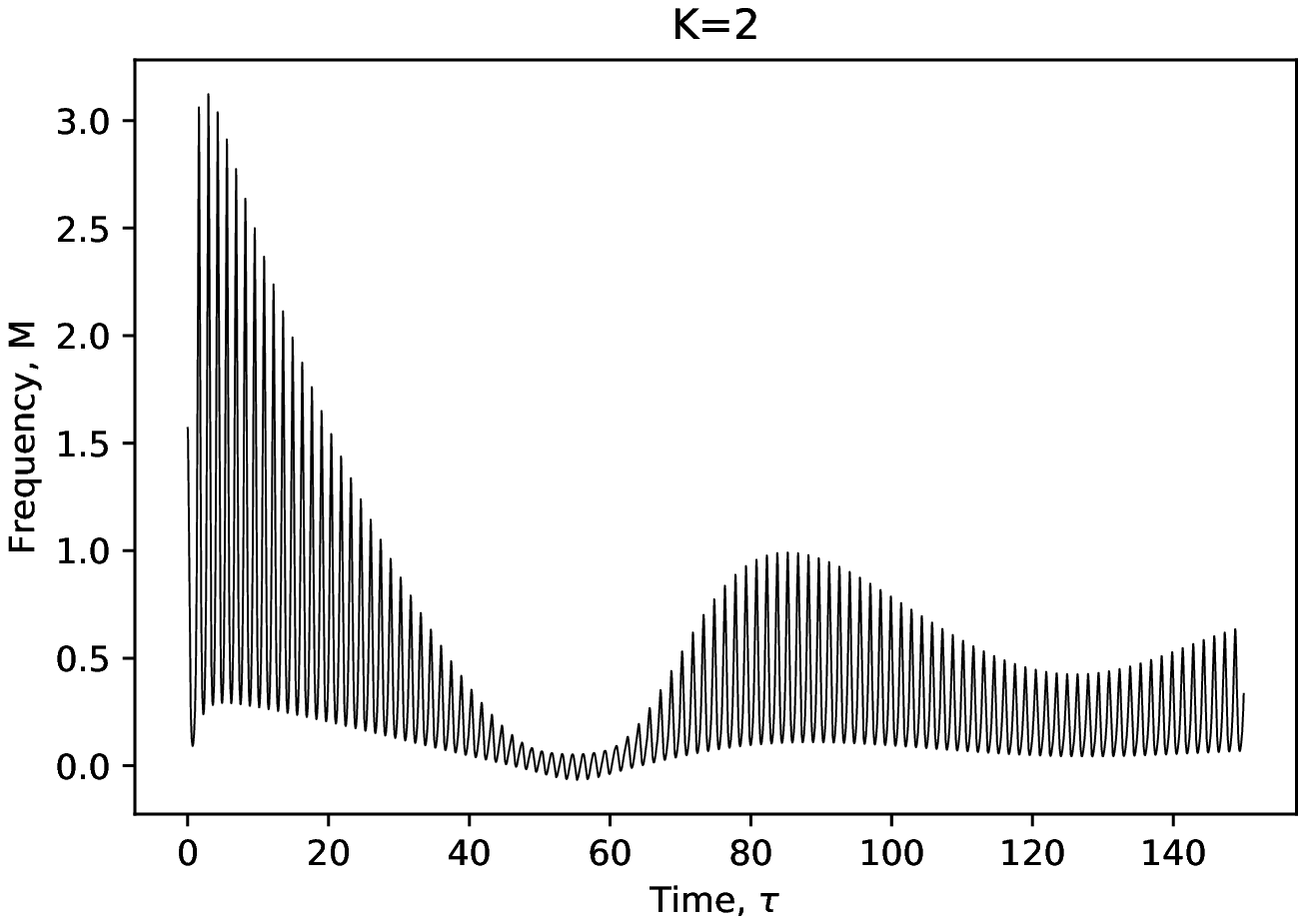} 
\end{minipage}%
\begin{minipage}{0.5\textwidth}
\includegraphics[width=3.in,height=2.in]{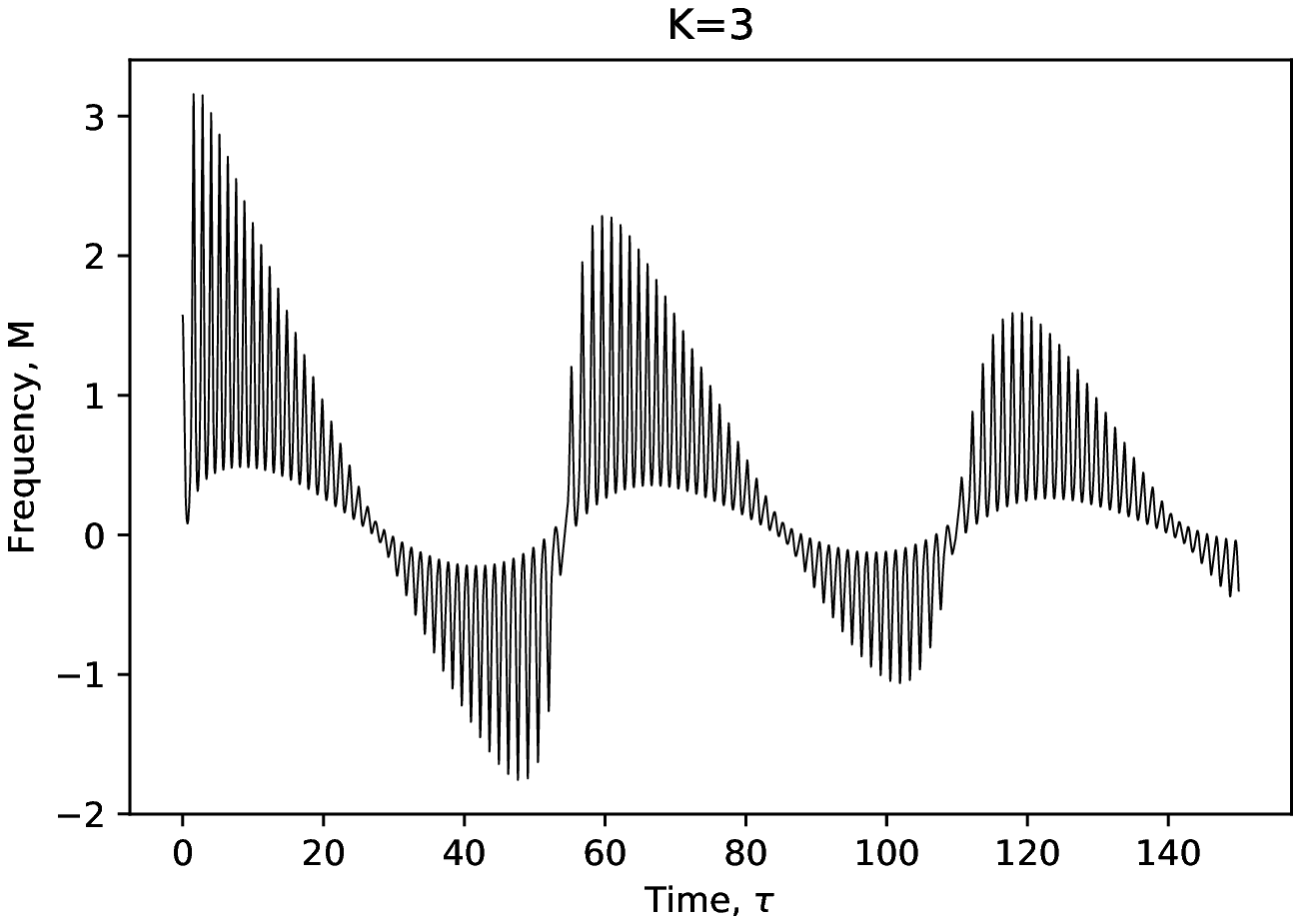} 
\end{minipage}\\
\begin{minipage}{0.5\textwidth}
\includegraphics[width=3.in,height=2.in]{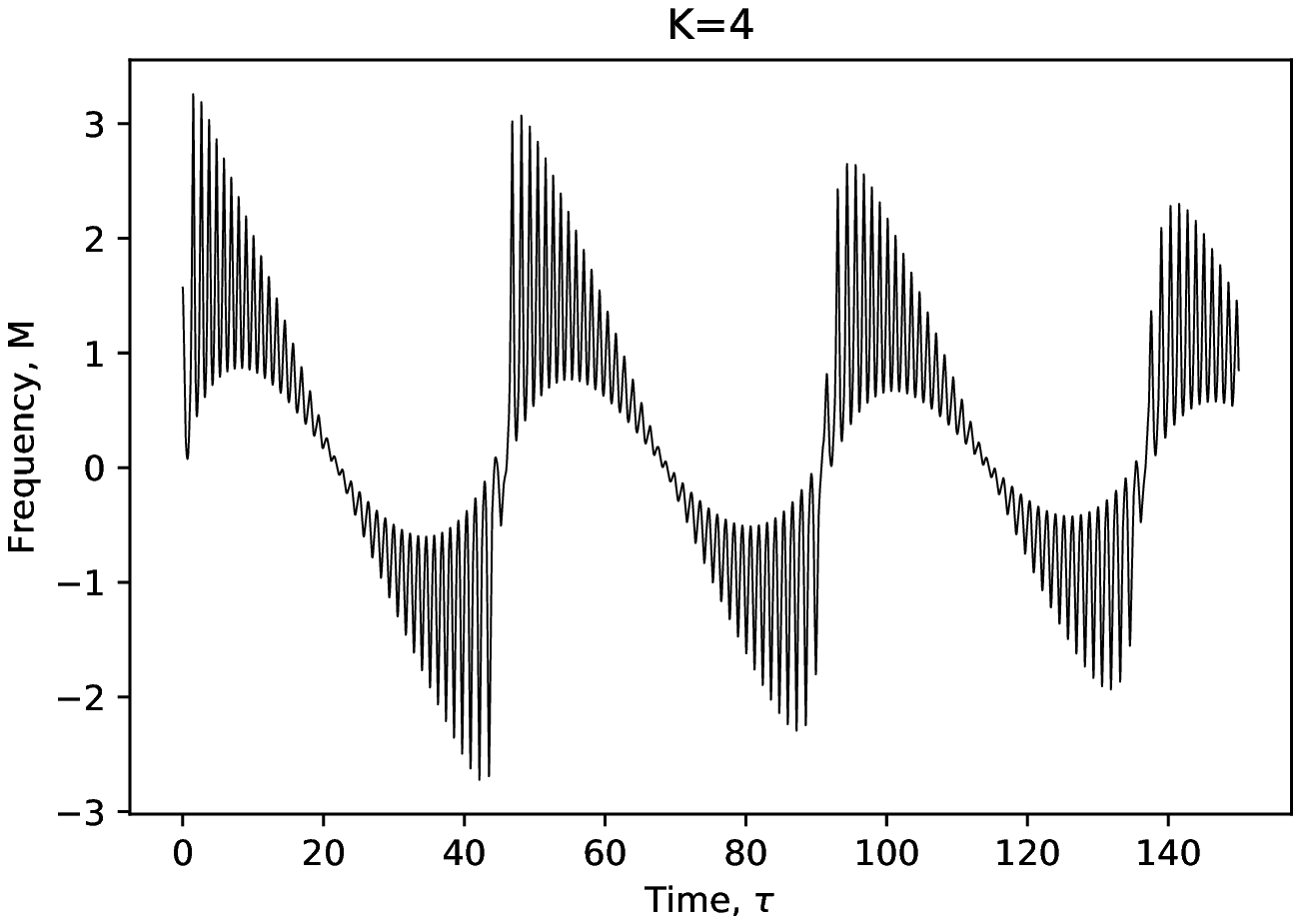} 
\end{minipage}%
\begin{minipage}{0.5\textwidth}
\includegraphics[width=3.in,height=2.in]{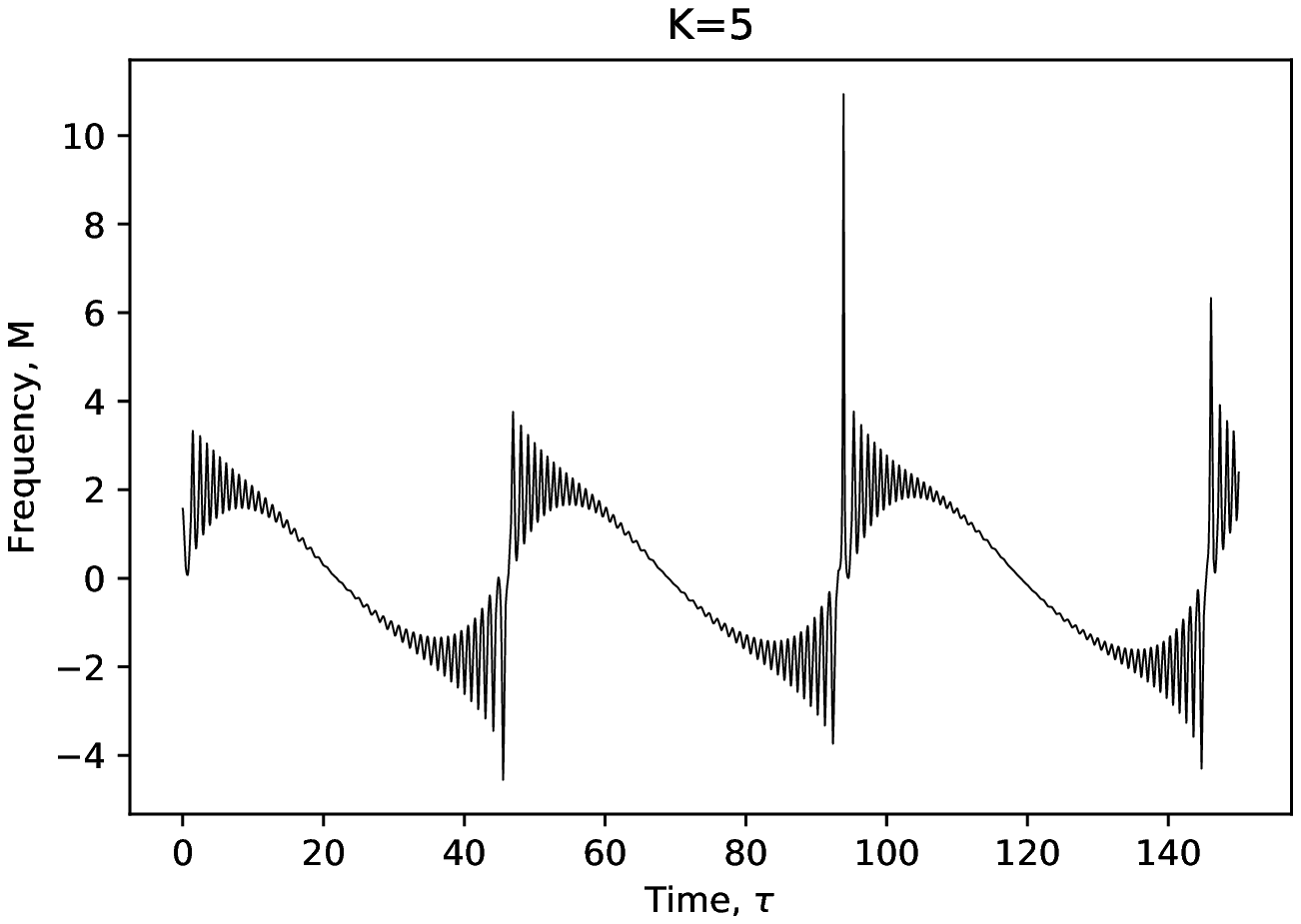}
\end{minipage}
\caption{\label{fig6} Temporal profile of the instantaneous frequency $M(\tau)$ of the laser, generated numerically for different values of $K$ and for $\eta= 0.25$.} 
 \end{figure}

\begin{figure}\centering
\begin{minipage}{0.5\textwidth}
\includegraphics[width=3.in,height=2.in]{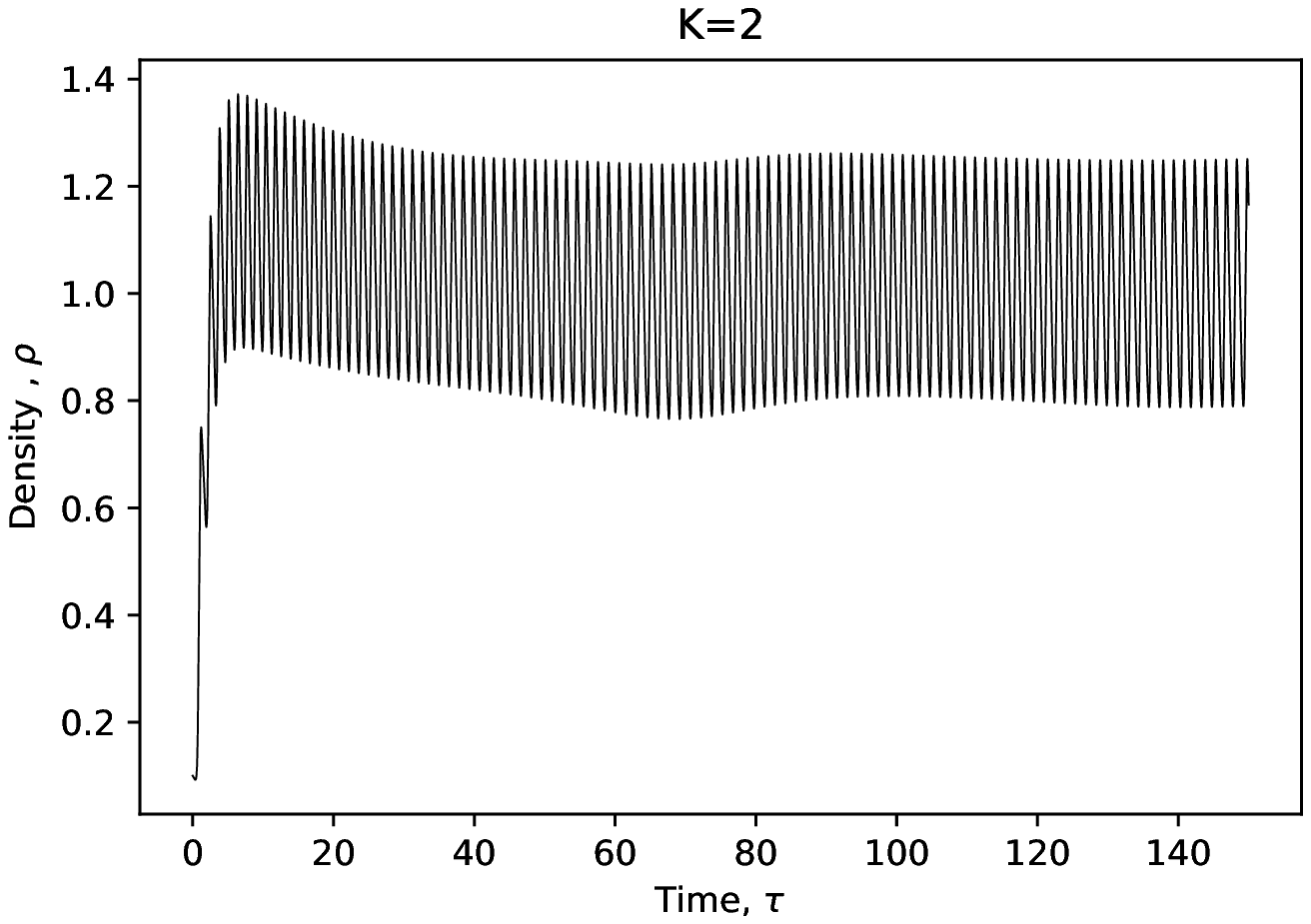} 
\end{minipage}%
\begin{minipage}{0.5\textwidth}
\includegraphics[width=3.in,height=2.in]{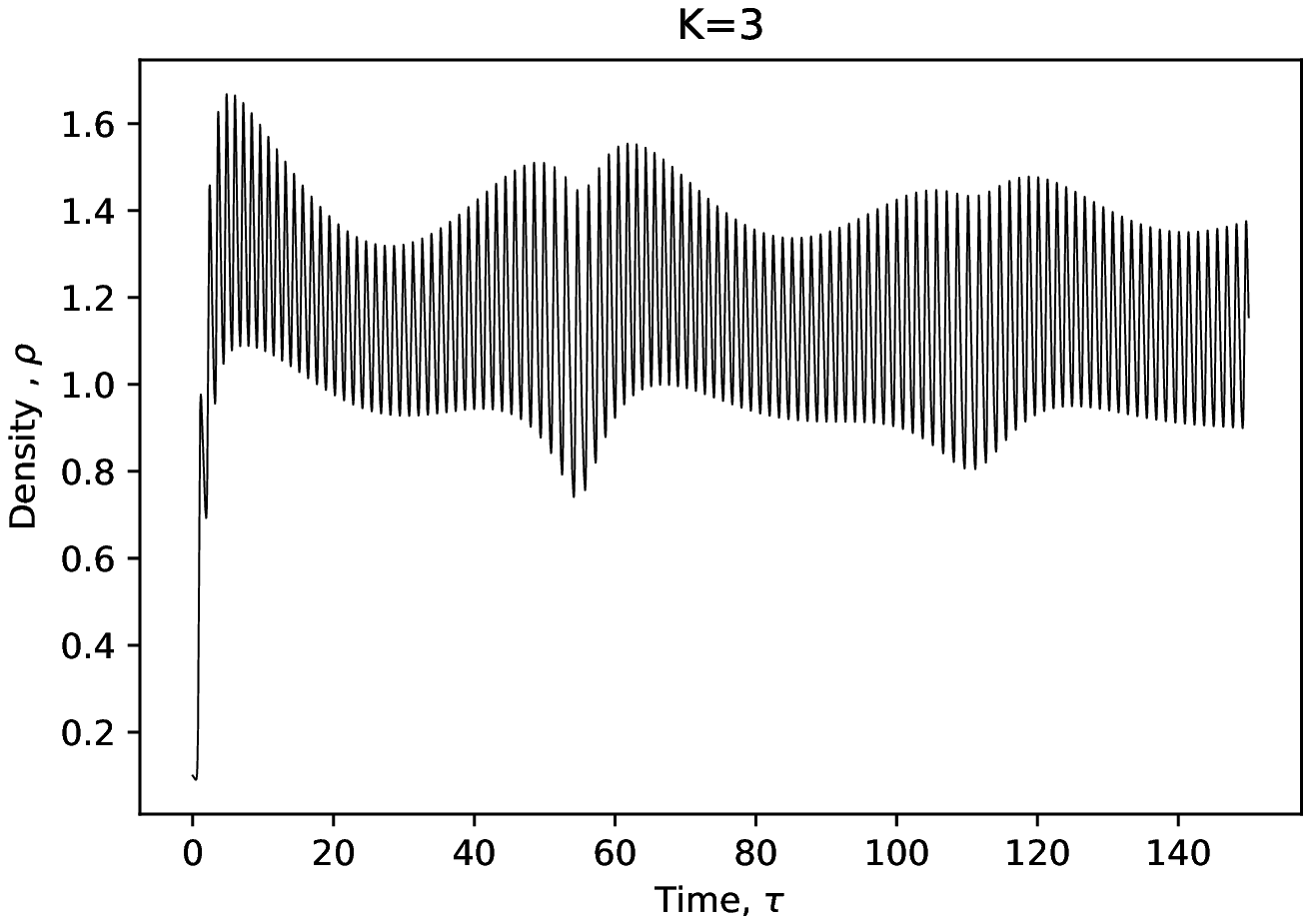} 
\end{minipage}\\
\begin{minipage}{0.5\textwidth}
\includegraphics[width=3.in,height=2.in]{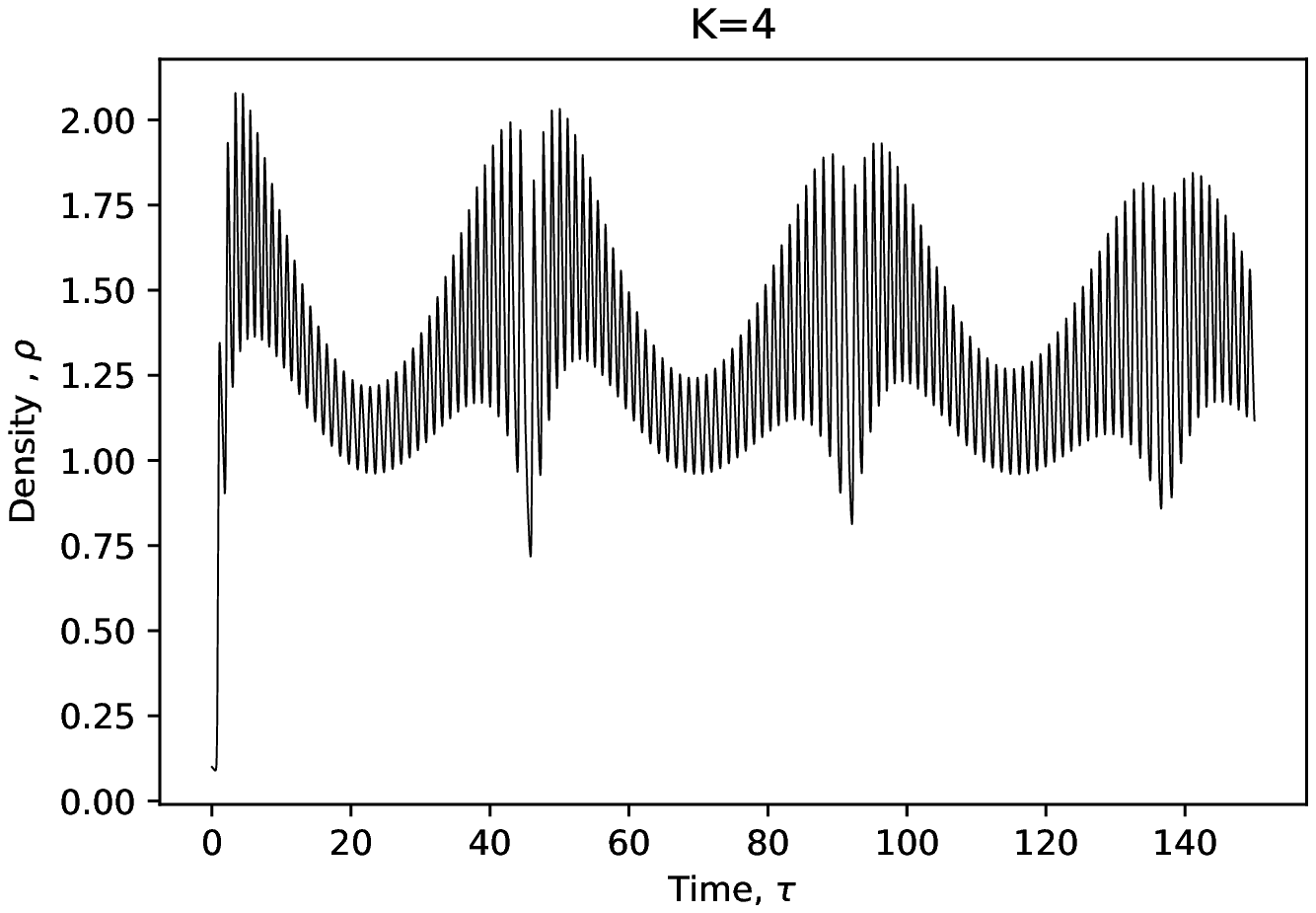} 
\end{minipage}%
\begin{minipage}{0.5\textwidth}
\includegraphics[width=3.in,height=2.in]{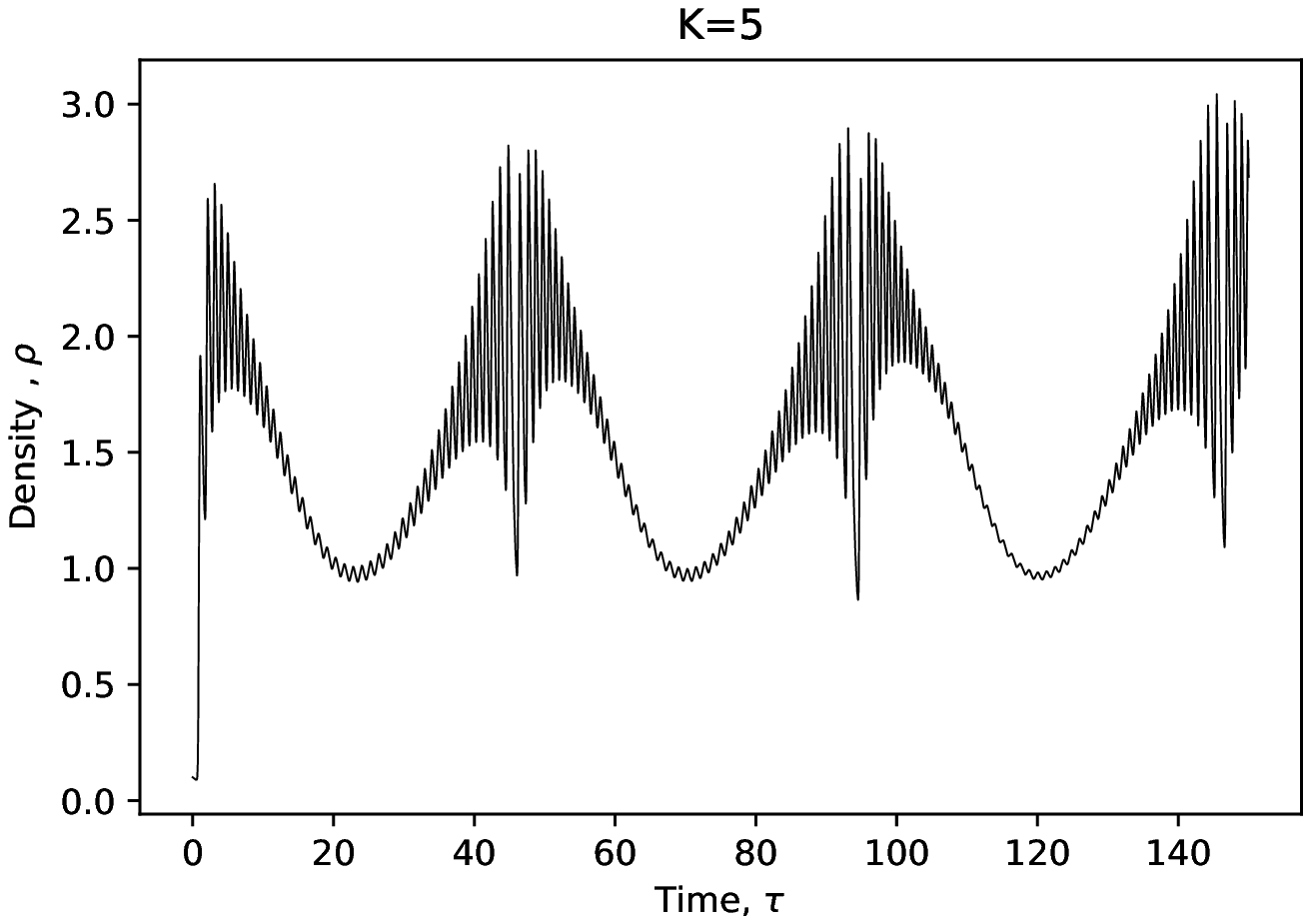}
\end{minipage}
\caption{\label{fig7} Temporal profile of the plasma density $\rho(\tau)$, generated numerically for different values of $K$ and for $\eta= 0.25$.} 
\end{figure}

 To highlight the influence of $\eta$ on numerical solutions, in figs (\ref{fig5}), (\ref{fig6}) and (\ref{fig7}), time series of $A(\tau)$, $M(\tau)$ and $\rho(\tau)$ are represented now for $\eta=0.25$ and keeping all other characteristic parameters fixed.  
Temporal profiles of the laser amplitude $A(\tau)$, the laser instantaneous frequency $M(\tau)$ and the plasma  density $\rho(\tau)$ for $\eta=0.25$, represented in figs. \ref{fig5}, \ref{fig6} and \ref{fig7} respectively, are very clearly different from results obtained when $\eta=0.005$: the exponential growth of $A(\tau)$ that characterized this quantity for small values of the electron diffusion coefficient $\eta$ and for $K=2$, is completely wiped out at higher values of $\eta$. In this later case, the laser amplitude $A(\tau)$ instead reproduces a regular train of pulse structures, whose temporal period (i.e. the pulse repetition rate in the periodic train) is increased as $K$ increases. So to say, temporal profiles of the laser amplitude obtained from numerical simulations suggest that stronger electron diffusion processes will favor trains of well separated pulses with large repetition rates. The plasma density in this case too displays a relatively more regular structure, as seen in fig. \ref{fig7}, compared with numerical results of the same quantity for smaller value of $\eta$ and represented in fig. \ref{fig4}. The laser instantaneous frequency $M(\tau)$ also becomes more regular as $\eta$ is increased, as seen in fig. \ref{fig6} compared with the behaviour observed in fig. \ref{fig3} when $\eta$ is relatively smaller. 
 
\section{Conclusion}
Nonlinear phenomena in artifical optical systems have attracted a great deal of interest over the past years, due to their contributions to innovative technologies in optics and photonics in general for all‐optical signal processings. These technologies rely on exotic optical circuits fabricated via different methods, including femtosecond laser inscriptions in optical and photonic materials \cite{hu}. Femtosecond laser inscription technique rests on a nonlinear absorption of an ultrashort laser pulse, focused in the bulk of a transparent material with eventually an optical Kerr nonlinearity \cite{r9}. Since the laser interaction with the transparent material is a nonlinear process \cite{r6,r7,r8,r9,r10,r11}, a localized modification of the bulk material is favored which generates a positive refractive index change \cite{r9,mar1,mar2,mar3,mar4}. \par To gain a better understanding of the laser dynamics interacting with transparent materials, recently some theoretical models were proposed \cite{r6,r7,r8,r9,r10,r11}. In these models it is assumed that upon absorption, the femtosecond laser propagates in the transparent material storing energy along its path. The energy stored causes an ionization of the bulk material and subsequently induces an electron plasma in the material. Theoretically, the femtosecond laser propagation in the optical Kerr medium can be represented by a cubic complex Ginzburg-Landau equation, with an additional term accounting for multiphoton absorption processes. The plasma density is assumed inhomogeneous, and its time evolution is governed by a first-order ordinary differential equation, where distinct terms can be present accounting for distinct physical processes assumed to contribute to the generation of the electron plasma. In this respect, in ref. \cite{r9} it was stressed that a physically acceptable model for the dynamics of plasma density should include simultaneously the avalanche ionization, electron-hole radiative recombination processes, the multiphoton ionization and electron diffusion processes. In the present work we examined the dynamics of this last model, laying emphasis on the influence of electron diffusion processes on temporal profiles of the amplitude and instantaneous frequency of the femtosecond laser, as well as of the plasma density. We found that electron diffusion processes favor pulse trains with relatively large pulse repetition rates, and weakly damped periodic anharmonic oscillations of the plasma density. The present study therefore clearly shows that electron diffusion processes play a relevant role in femtosecond laser inscriptions in transparent media with Kerr nonlinearity.   

\section*{Acknowledgments}
E. O. Akeweje wishes to thank AIMS-Ghana for the opportunity of visit, and E. Opoku Gyasi for enriching discussions during his stay at AIMS Ghana.

\section*{Disclosure statement}

The authors declare no potential conflict of interest.

\end{document}